%% file: main.tex
\documentclass[10pt, journal, letterpaper]{IEEEtran}
\IEEEoverridecommandlockouts
\usepackage{import}
\usepackage{graphicx}
\usepackage{amsfonts}
\usepackage{amssymb}
\usepackage{amsmath}
\usepackage{tabularx}
\usepackage{multirow, makecell}
\usepackage[dvipsnames]{xcolor}
\graphicspath{ {./figures/} }
\usepackage{subcaption}
\usepackage{multirow}
\usepackage{textcomp}
\usepackage{algorithm}
\usepackage{algorithmic}
\DeclareMathOperator*{\argmax}{argmax}
\usepackage[style=base]{caption}

\ifCLASSOPTIONcompsoc
  \usepackage[nocompress]{cite}
\else
  \usepackage{cite}
\fi

\begin{document}
\title{Scheduling Optimization of Heterogeneous Services by Resolving Conflicts}

% \title{Conflict_based 5G New Radio Resource Allocation for Heterogeneous Services}

% \title{Resource Allocation Optimization with Mixed Numerology and Time Scaling for Heterogeneous Services}

% \title{Conflict_based 5G New Radio Resource Allocation for Heterogeneous Services}

% \title{ Radio Resource Allocation with Mixed Numerology and Time Scaling for Heterogeneous Services}

% \title{ Multi-Numerology Radio Resource Scheduling with Flexible Time Scaling for Heterogeneous Services}

\author{Sotiris Skaperas, \textit{Member, IEEE}, Nasim Ferdosian, \textit{Member, IEEE}, Arsenia Chorti, \textit{Senior Member, IEEE} and Lefteris Mamatas, \textit{Member, IEEE}% <-this % stops a space

\thanks{S. Skaperas and L. Mamatas are with Dep. of Applied Informatics, University of Macedonia, Thessaloniki, Greece, e-mails: \{sotskap, emamatas \}@uom.edu.gr}% <-this % stops a space

\thanks{N. Ferdosian and A. Chorti are with ETIS UMR 8051,
CY-Tech, ENSEA, France, e-mails: \{nasim.ferdosian, arsenia.chorti \}@ensea.fr}% <-this % stops a space
}

\maketitle
\begin{abstract}
Fifth generation (5G) new radio introduced flexible numerology to provide the necessary flexibility for accommodating heterogeneous services. However, optimizing the scheduling of heterogeneous services with differing delay and throughput requirements
over 5G new radio is a challenging task. In this paper, we investigate near optimal, low complexity scheduling of radio resources for ultra-reliable low-latency communications (URLLC) when coexisting with enhanced mobile broadband (eMBB) services. We demonstrate that maximizing the sum throughput of eMBB services while servicing URLLC users, is, in the long-term, equivalent to minimizing the number of URLLC placements in the time-frequency grid; this result stems from reducing the number of infeasible placements for eMBB,  to which we refer to as ``conflicts’’. To meet this new objective, we propose and investigate new conflict-aware heuristics; a family of ``greedy’’ and a lightweight heuristic inspired by bin packing optimization, all of near optimal performance. Moreover, having shed light on the impact of conflict in layer-2 scheduling, non-orthogonal multiple access (NOMA) emerges as a competitive approach for conflict resolution, in addition to the well established increased spectral efficiency with respect to OMA. The superior performance of NOMA, thanks to alleviating conflicts, is showcased by extensive numerical results.
\end{abstract}
\IEEEpeerreviewmaketitle

 \section{Introduction}\label{Intro}
 \import{Sections/}{introduction}

\section{Problem Formulation}\label{Prob}
We first provide a review of basic concepts in 5G NR flexible numerology and detail the considered scheduling problem.

\subsection{Background on 5G NR flexible numerology}
5G NR Release-15 \cite{3GPP} defines a flexible numerology with subcarrier spacing (SCS) of $15$, $30$, and $60$ kHz below $6$ GHz, and $60$ and $120$ kHz above $6$ GHz, compared to long-term evolution (LTE) which uses a fixed numerology with SCS of $15$ kHz below $6$ GHz. 5G NR also defines a $10$ msec frame, with each frame divided into $10$ subframes of $1$ msec, which are further divided into one or more mini-slots. A mini-slot comprises $14$ OFDM symbols for a configuration using normal cyclic prefix, or $12$ OFDM symbols for extended cyclic prefix. 

 In 5G NR, the mini-slot size is defined according to the symbol duration, which is inverse to the SCS, to ensure the orthogonality of the subcarriers. By using higher SCS, the symbol duration decreases and hence also the mini-slot size, which is beneficial for lower latency \cite{semiari2019integrated}. URLLC traffic requires extremely low delays, often lower than 1 ms \cite{sutton2019enabling}. The URLLC latency requirements can only be satisfied if the transmission duration and round-trip-time (RTT) are shorter than the corresponding latency constraint.
 The major challenges related to radio resource optimization for URLLC systems are described in \cite{she2017radio}.% It is shown that, by adopting the new short slot structures, the URLLC resource management becomes more flexible. The URLLC latency requirement can be satisfied if its transmission duration and RTT are shorter than its latency constraint. Authors in \textcolor{red}{\cite{}} investigate the functionality of the envisioned NR radio interfaces to provide URLLC services. It is revealed that NR meets the ITU 5G requirements on URLLC services, at the cost of reduced spectral efficiency for mobile broadband services. \cite{marijanovic2019multi}.

\subsection{Scheduling problem formulation}
\import{Sections/}{problem}

\section{Heuristic Algorithms for Conflict Resolution}\label{Alg}
\import{Sections/}{algorithm}

\begin{figure*}[!ht]\label{fixed_flexible}
    \centering
    \begin{subfigure}[b]{0.48\textwidth}
         \centering
         \includegraphics[width=\textwidth]{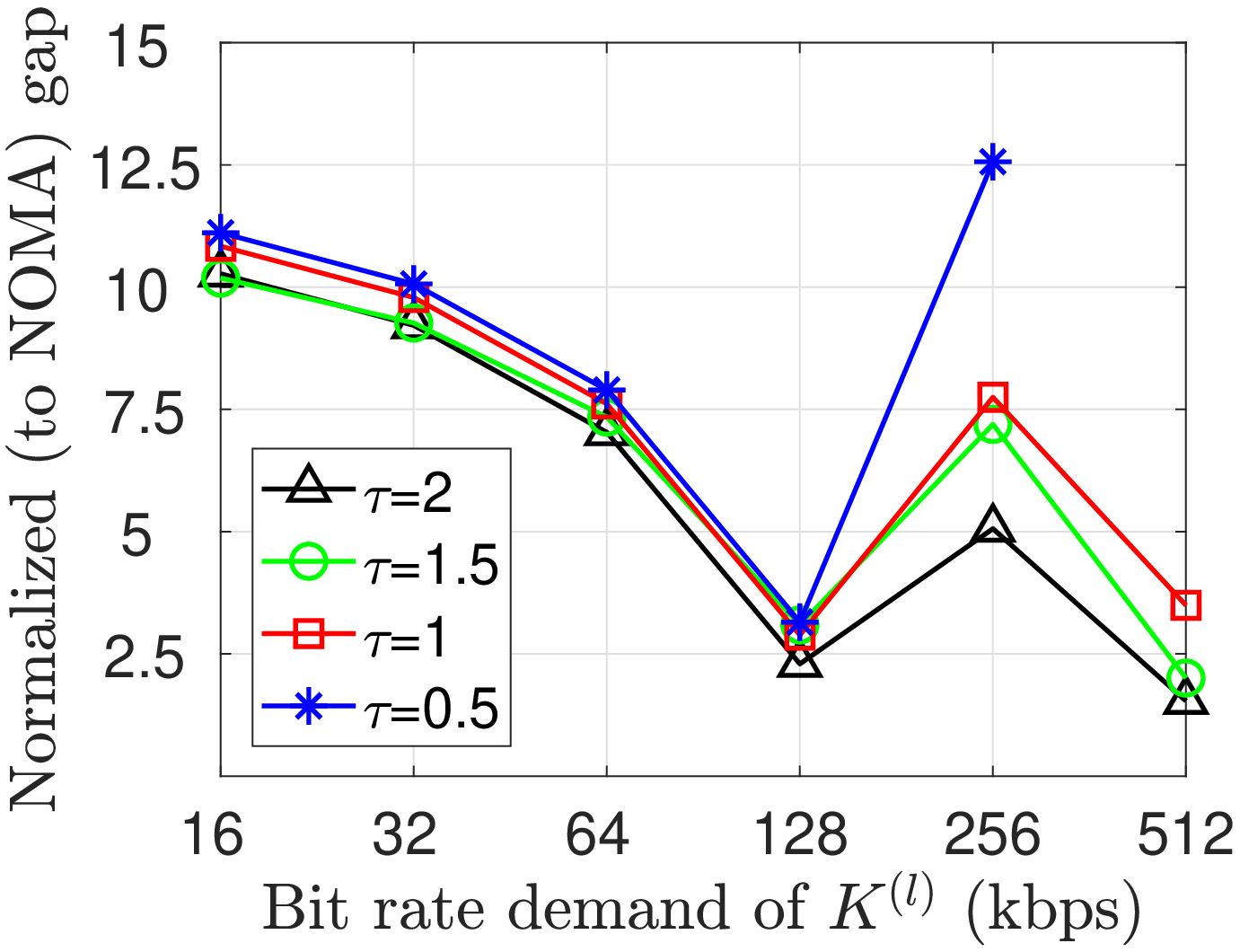}
         \caption{Flexible}
    \end{subfigure}%
    \begin{subfigure}[b]{0.48\textwidth}
         \centering
         \includegraphics[width=\textwidth]{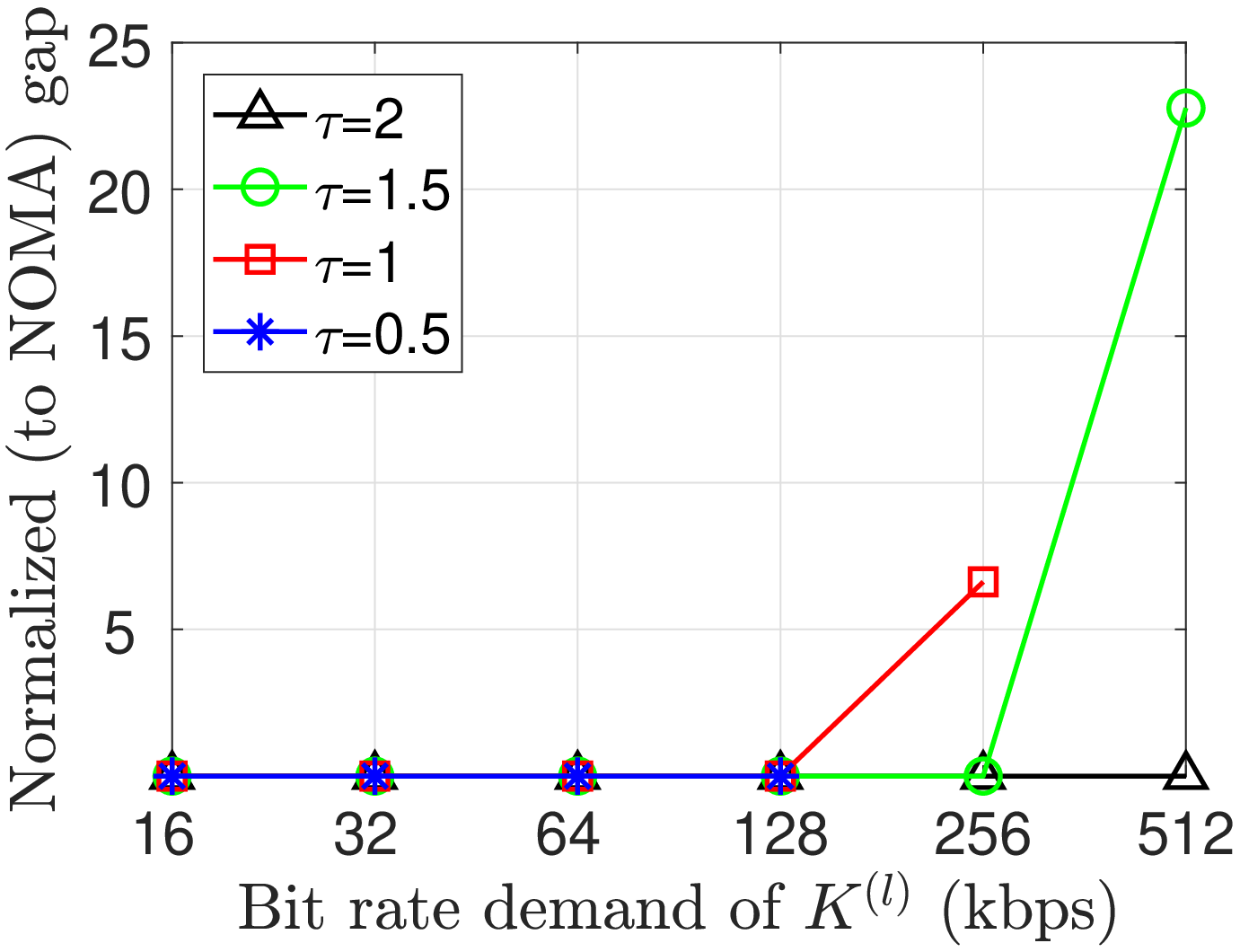}
         \caption{Multiple-fixed}
     \end{subfigure}
          \begin{subfigure}[b]{0.32\textwidth}
         \centering
         \includegraphics[width=\textwidth]{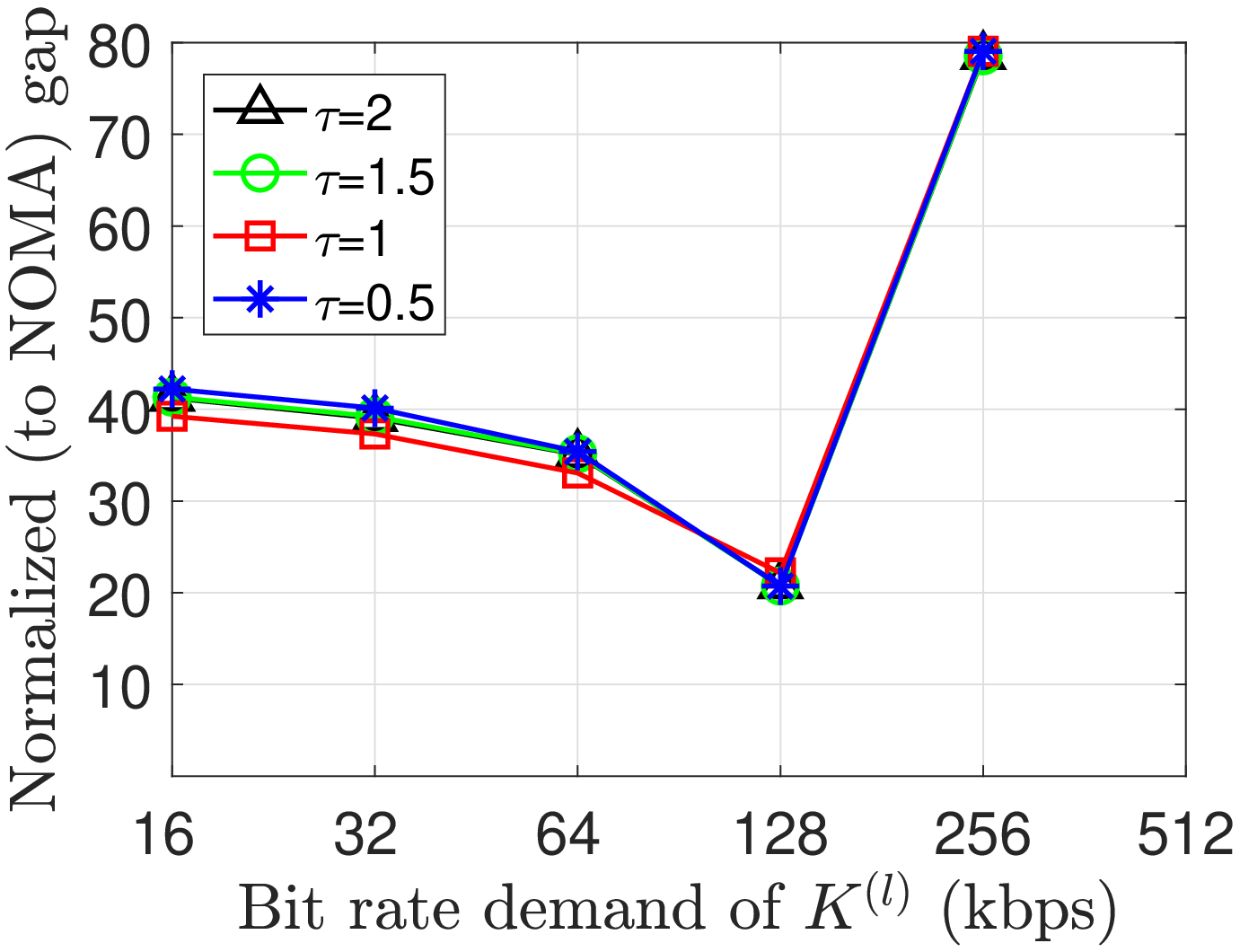}
         \caption{Shape 1}
     \end{subfigure}
     \begin{subfigure}[b]{0.32\textwidth}
         \centering
         \includegraphics[width=\textwidth]{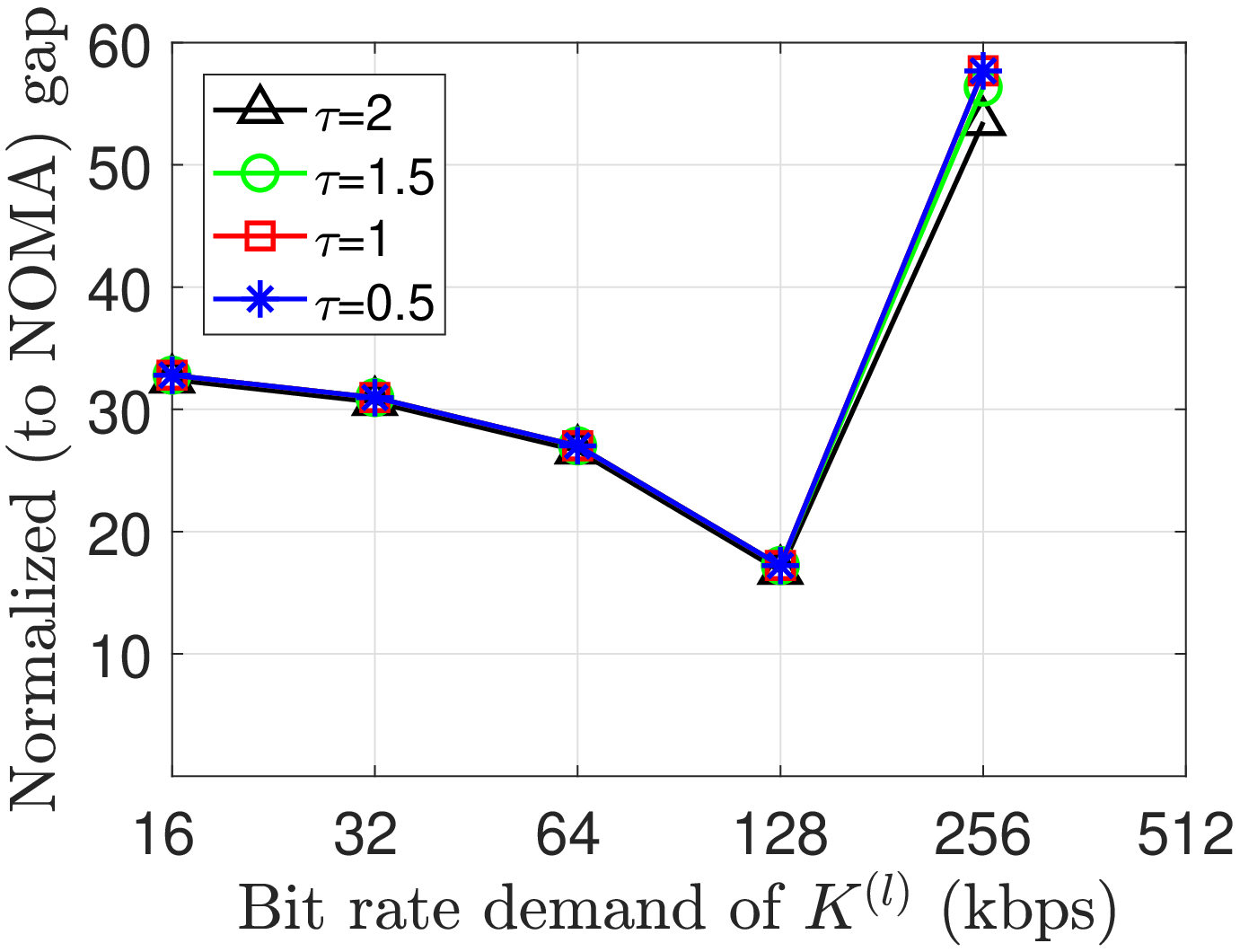}
         \caption{Shape 2}
     \end{subfigure}
     \begin{subfigure}[b]{0.32\textwidth}
         \centering
         \includegraphics[width=\textwidth]{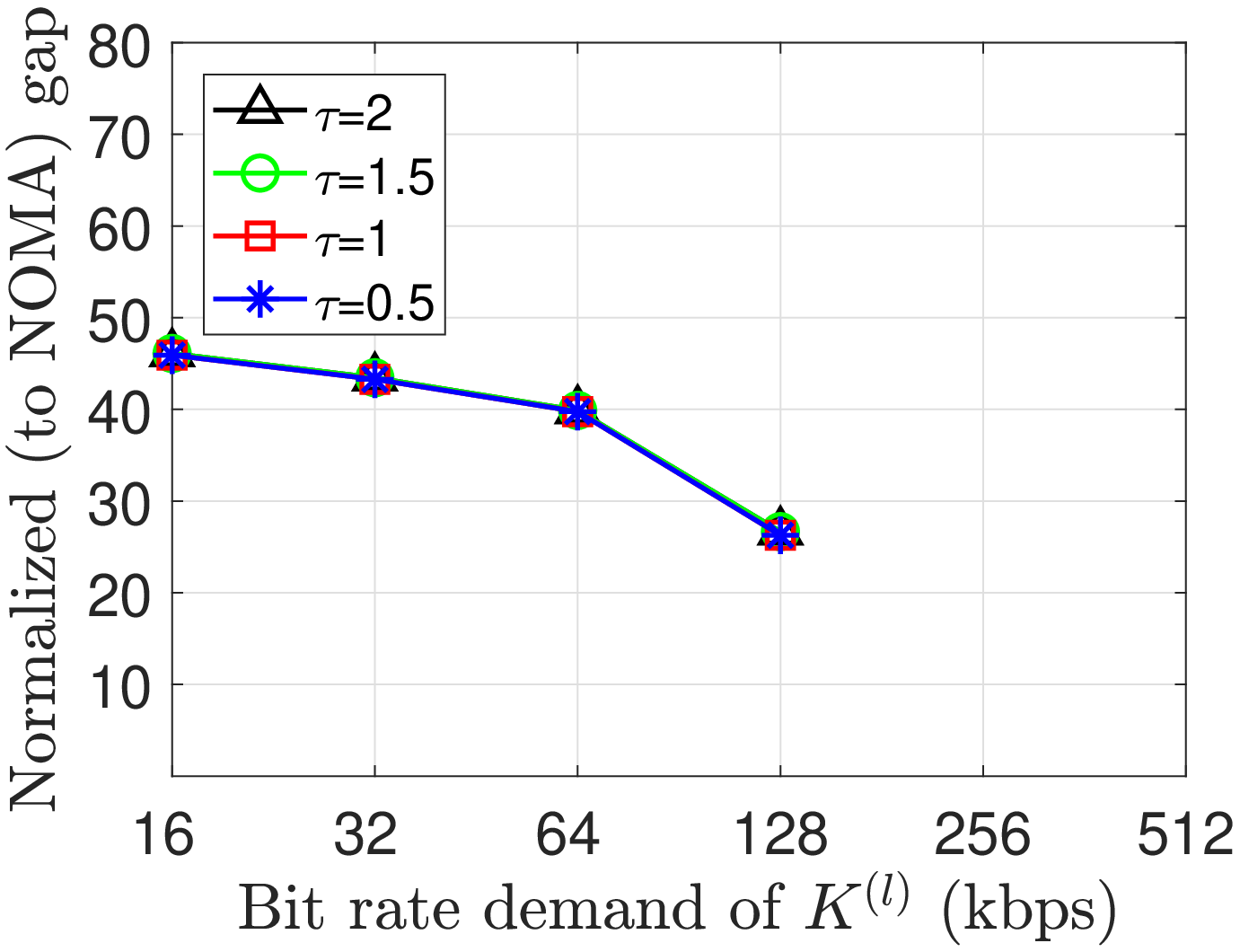}
         \caption{Fixed (Shape 2) structure.}
         
     \end{subfigure}
     \caption{Normalized (to NOMA) gap of the sum bit rate of the $\mathcal{K}^{(c)}$ services between NOMA and OMA schemes. The y-axes measure percentages. Non existing values indicate infeasible solutions.} %We exclude the delay tolerance value $\tau_k=0.25$ msec, $k\in\mathcal{K}^{(\ell)}$ from fixed and multiple-fixed numerology results, since they provide infeasible solutions for both OMA and NOMA schemes.}
    \label{fig:gaps}
    \end{figure*}

\section{NOMA for Downlink Scheduling}
 In this section, we re-examine P0 under the assumption that it is possible to employ NOMA in the downlink to schedule different services, even at the mini-slot level \cite{popovski20185g}. %considering both flexible and non flexible numerology. 
 In contrast to the scheduling optimization problem as formulated in P0, NOMA allows overlapping amongst the blocks, either full or partial (of some mini-slots). In light of this, P0 is reduced to a linear programming (LP) problem that we refer to as P1, in which the optimization parameter is now a real number $x_{b,k}\in[0,1]$, 
 \begin{align}
\text{[P1]} \quad \max_{x_{b,k} \in [0,1]} &  \sum_{b\in\mathcal{B}}\sum_{k\in\mathcal{K}^{(c)}} r_{b,k}x_{b,k}, \\
%& R_{total}-\sum_{b\in\mathcal{B}}\sum_{k\in\mathcal{K}^{(l)}} r_{b,k}x_{b,k}=\nonumber\\
%&R_{total}-\sum_{b\in\mathcal{B}}\sum_{k\in\mathcal{K}} cf_{b, k}r_{b,k}x_{b,k}\\
\text{s.t.} \quad & \sum_{b\in\mathcal{B}}r_{b,k}x_{b,k}\geq q_{k}, \quad k\in\mathcal{K}^{(\ell)} ,\\
           & \sum_{b\in\mathcal{B}}\sum_{k\in\mathcal{K}}a_{b,i}x_{b,k}\leq \tilde{r}, \quad i\in\mathcal{I},
\end{align}
where $\tilde{r}$ denotes the  normalized sum throughput per block achieved with NOMA, which in previous works has been shown to be superior to OMA (i.e., greater than unity); note that in P0, constraint (\ref{eq:constraint2}) is upper bounded to unity.  This points out a further gain in using NOMA due to the increase in per resource block utilization. However, as in this work we aim primarily at demonstrating the gains brought about due to conflict avoidance, in the numerical results presented in Section V we simply use $\tilde{r}=1$. 
    
\begin{figure}[t]
     \centering
     \begin{subfigure}[b]{0.23\textwidth}
         \centering
         \includegraphics[width=\textwidth]{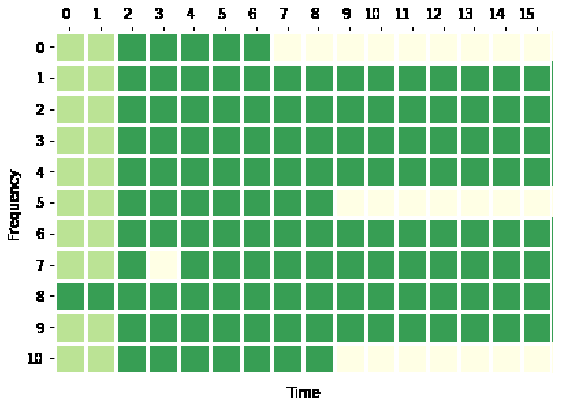}
         \caption{OMA case, $\tau_{k}=0.5$ ms and $q_{k}=32$ kbps.}
     \end{subfigure}
     \quad
     \begin{subfigure}[b]{0.23\textwidth}
         \centering
         \includegraphics[width=\textwidth]{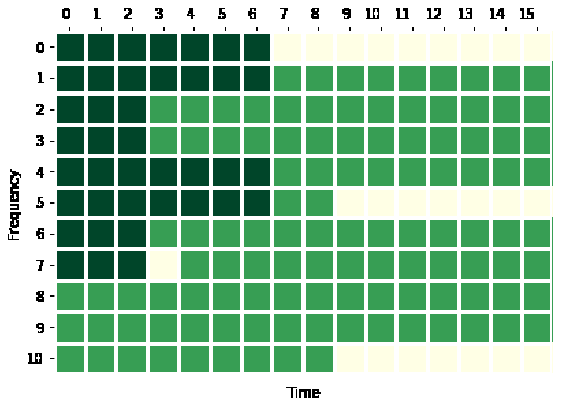}
         \caption{NOMA case, $\tau_{k}=0.5$ ms and $q_{k}=32$ kbps.}
     \end{subfigure}
          \begin{subfigure}[b]{0.23\textwidth}
         \centering
         \includegraphics[width=\textwidth]{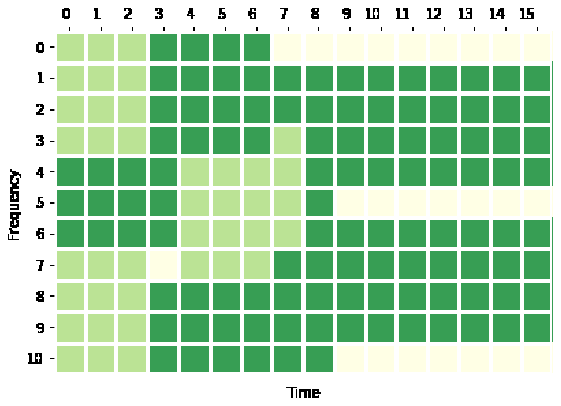}
         \caption{OMA case, $\tau_{k}=1$ ms and $q_{k}=256$ kbps.}
     \end{subfigure}
     \quad
     \begin{subfigure}[b]{0.23\textwidth}
         \centering
         \includegraphics[width=\textwidth]{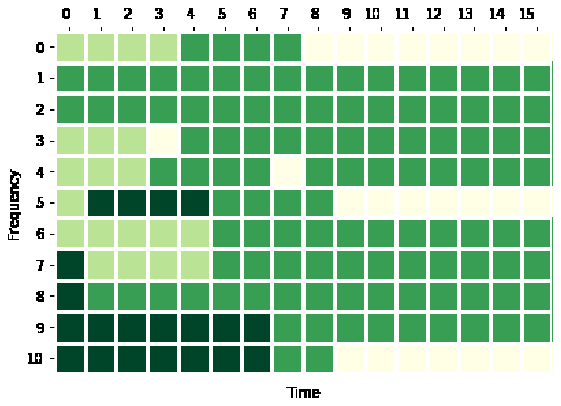}
         \caption{NOMA case, $\tau_{k}=1$ ms and $q_{k}=256$ kbps.}
     \end{subfigure}
    \begin{subfigure}[b]{0.23\textwidth}
         \centering
         \includegraphics[width=\textwidth]{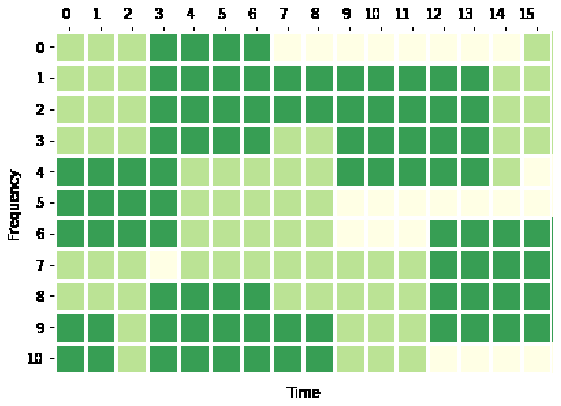}
         \caption{OMA case, $\tau_{k}=2$ ms and $q_{k}=512$ kbps.}
     \end{subfigure}
     \quad
     \begin{subfigure}[b]{0.23\textwidth}
         \centering
         \includegraphics[width=\textwidth]{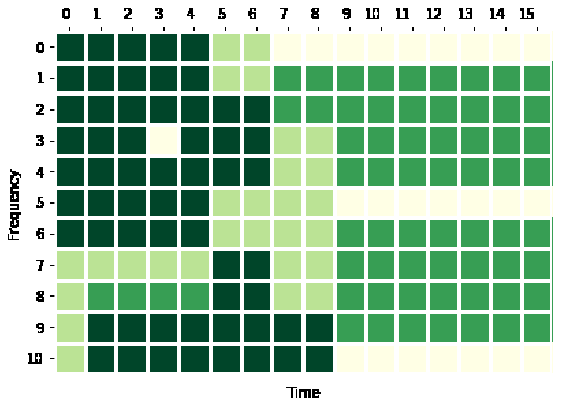}
         \caption{NOMA case, $\tau_{k}=2$ ms and $q_{k}=512$ kbps.}
     \end{subfigure}
\caption{Resource allocation of URLLC (light green) and eMBB (green) services, for OMA (first column) and NOMA (second column). Light yellow denotes zero throughput mini-slots. Dark green denotes overlapping of mini-slots thanks to using NOMA.}
\label{fig:pyt}
\end{figure}  
 
P1 can be efficiently solved (optimal solution) by using the simplex method, interior point methods, or the ellipsoid method \cite{Boyd}, with respect to the infeasible solutions. In the specific problem, it is preferable to solve the dual instead of the primal problem, since the computation time increases much more rapidly with the number of constraints than with the number of variables. Moreover, the ellipsoid and the
interior point methods are mathematically iterative and need significantly more computing
resources than does the simplex algorithm for small linear programming problems. Hence, we employ the dual simplex algorithm, for which in the worst case scenario with an exponential number of corners, an
exponential number of steps can be taken to find the optimal corner.

\section{Numerical Results}\label{Resu}
\import{Sections/}{results}

\section{Conclusion}\label{Con}
\import{Sections/}{conclusion}

\section*{Acknowledgements}
We thank the authors of \cite{you2018resource} for kindly sharing their simulation codes in IEEE DataPort \cite{ch8e_x385_18}. N. Ferdosian and A. Chorti have been supported by the INEX project eNiGMA of the Inititative of Excellence of CYU.

\bibliographystyle{IEEEtran}
\bibliography{biblio.bib}
\end{document}

%% file: Sections/introduction.tex
The International telecommunication union (ITU) has defined new requirements and capabilities on 5G mobile communication systems to support a wide variety of new devices and services with diverse quality of service (QoS) requirements and characteristics \cite{ITU}. The 3rd generation partnership project (3GPP) standardized 5G in the form of a novel radio interface technology, referred to as new radio (NR) \cite{3GPP}. 5G NR introduced flexible numerology and frame structure  to accommodate heterogeneous service requirements, by supporting various values of subcarrier spacing and symbol / frame duration. Optimizing resource allocation in the NR numerology setting to deliver heterogeneous QoS requirements remains a challenging task \cite{sadi2020flexible, pocovi2018multiplexing, akhtar2018downlink, marijanovic2019multi, soldani20185g}.

In 5G and beyond, ultra-reliable low-latency communication (URLLC) services with extreme delay constraints will coexist with enhanced mobile broadband (eMBB) \cite{semiari2019integrated}, that require very high bit rates (Gigabits per second) and have moderate latency (a few milliseconds) requirements \cite{chen2018ultra}. Moreover, at present, URLLC services are expected to have lower traffic volumes than eMBB services \cite{sachs20185g}, while this might not hold in the future for applications such as virtual reality and haptics. In this framework, the design of radio resource allocation strategies for URLLC traffic when coexisting with eMBB has been a focal point of recent research efforts \cite{zhang2020dynamic, pocovi2018joint, korrai2020ran, korrai2019slicing}.

% From Thesis
In this direction, two approaches have been adopted by the 3GPP. The first is based on a ``puncturing'' framework: according to this, eMBB traffic is scheduled initially at the beginning of the slots; upon arrival of URLLC traffic, the latter is being prioritized and dynamically overlapped at mini-slots of ongoing eMBB transmissions (which are punctured, i.e., dropped). In the second approach, known as preemptive scheduling, resources are preemptively reserved for URLLC, before the demands are placed\cite{you2018resource, esswie2019preemption, esswie2018opportunistic, pocovi2018achieving}.

Based on puncturing scheduling, the studies in \cite{anand2020joint, pradhan2020joint, pedersen2017punctured, alsenwi2019embb}  considered resource allocation strategies for the coexistence of URLLC and eMBB. The authors in \cite{anand2020joint} consider three types of models - threshold, linear and convex -  to describe the eMBB data rate loss associated with the incoming URLLC traffic. Furthermore the authors in \cite{pedersen2017punctured} propose a punctured scheduling approach for transmission of low latency communication (LLC) traffic multiplexed on a shared channel with eMBB. Another approach is proposed in \cite{alsenwi2019embb}, where a risk-sensitive model was introduced in order to ensure URLLC allocation but also to minimize losses for eMBB users. However, these strategies can result in significant losses in terms of data rates for eMBB services \cite{li2020deep} and may impact eMBB transmission reliability \cite{alsenwi2020intelligent}. 
% In \cite{anand2020joint}, \cite{pradhan2020joint} resource allocation strategies for the coexistence of URLLC and eMBB were proposed based on a ``puncturing'' framework: according to this, eMBB traffic was scheduled initially at the beginning of the slots; upon arrival of URLLC traffic, the latter was prioritized and dynamically overlapped at mini-slots of  ongoing eMBB transmissions, which were punctured. These approaches have been shown to result in significant losses in terms of data rates for eMBB services \cite{li2020deep}.  
Alternatively, the authors in \cite{you2018resource} studied the resource allocation of eMBB and URLLC services by preemptively reserving resources for URLLC. 
Such solutions ensure advantageous conditions for URLLC packets when they are generated, at the cost of wasting resources in absence of URLLC transmissions \cite{alsenwi2020intelligent}.

A flexible numerology and frame structure was explicitly considered in \cite{you2018resource} by defining a time-frequency resource grid, containing different types of resource blocks of different shapes, expanding over different time spans and frequency ranges. Exploiting this flexibility to optimize the resource allocation to different services while ensuring their QoS requirements, was shown to be an ${NP}$-hard problem. The resource allocation optimization over flexible numerology and frame structure while avoiding the assignment of overlapping blocks that will cause collision (i.e., puncturing), still remains a challenging task.

In this paper, we consider a flexible, 2-dimensional grid of resource blocks with different sizes in the time and frequency domains. The problem of identifying the resource allocation that maximizes the eMBB sum-rate is studied under the constraint of covering all URLLC throughput demands under different latency constraints ranging from $0.25$ to $2$ milliseconds (msec). We notice that due to the potential full or partial overlap of resource blocks, not all service placements are feasible once a specific placement has been executed. Here, we argue that managing infeasible placements is key and can be incorporated in low complexity algorithms explicitly. The main contributions of this work are outlined below:
\begin{enumerate}
    \item 
 We first re-formulate the problem of eMBB throughput maximization, introducing the URLLC conflicts minimization in the objective function. The novel concept of ``conflict" captures the penalties occurring due to the fact that orthogonal multiple access (OMA) does not allow overlapping of resources; as a result, OMA scheduling incurs a large number of infeasible resource allocation combinations. %, that we denote in the present as {conflicts}. 
 This new view angle allows for proposing completely novel solutions for the problem at hand. 
 
 \item Next, we propose three conflict-aware, multi numerology radio resource allocation heuristics to maximize scheduling efficiency for URLLC, when coexisting with eMBB services.  %\textcolor{red}{Lefteri, please revise this sentence if necessary}
 Three different functions of the i) average,  ii) the instantaneous (placement specific), or iii) the aggregate conflict are used to normalize the throughput utility function and incorporate penalties, when increasing conflicts. We argue and showcase through extensive simulation results that employing the proposed utilities improves the performance of proposed algorithms in the literature, as this in \cite{you2018resource}. 

% In this work, we first propose a conflict-aware, multi numerology radio resource allocation algorithm to maximize scheduling efficiency for URLLC when coexisting with eMBB services. The proposed scheduling approach results from re-formulating the standard eMBB throughput maximization problem in an equivalent form in which the objective is to minimize conflicts with URLLC in terms of resource allocation. This new problem is shown to be solved by jointly minimizing the placements of URLLC services in the time-frequency resource grid and the aggregate conflict, which can be treated as a specific instance of bin packing optimization. Simulation results show that a heuristic scheduling algorithm of linear complexity, provides a quick, lightweight and near optimal solution to resource allocation scheduling in URLLC and eMBB coexistence. 

\item Subsequently, we depart on a completely different approach with a high accuracy and low computational complexity. We treat the scheduling problem as a specific instance of bin packing optimization, solved by minimizing the placements of URLLC services in the time-frequency resource grid; to this end, we propose to group the resource blocks in different categories with respect to URLLC demands. Within each category, we solve a knapsack maximization of the sum eMBB throughput. Our proposal builds on previous results in \cite{MiroEurasip} and is inspired by the refined-first-fit family of heuristics to solve bin packing problems. Simulation results show that the novel heuristic algorithm, of complexity $N\log(N)$, provides a quick, lightweight and near optimal solution to the resource allocation scheduling of URLLC, when coexisting with eMBB. 

% Moreover, having shed light to the importance of minimizing conflicts between different services, the  utilization of NOMA schemes naturally emerges as a competitive candidate \cite{popovski20185g}. NOMA allows for the superposition of services, even at the 5G NR mini-slot level, by employing superposition coding at the transmitter and successive interference cancellation at the receivers \cite{song2017resource}, \cite{islam2016power}. NOMA has in the past been proposed as a competitive scheme to enhance throughput per resource block \cite{abusabah2018noma}; in the present work we provide further motivation for it's employment in B5G as the means to mitigate conflicts in the allocation of resource blocks, i.e., in layer $2$ scheduling. We provide an extensive set of numerical results that show the significant gains in terms of eMBB throughput when adopting NOMA in both fixed and flexible numerology settings.
\item
Moreover, having shed light to the importance of minimizing conflicts between different services, the  utilization of NOMA schemes \cite{abusabah2018noma}, \cite{popovski20185g} naturally emerges as a competitive candidate for interference management. NOMA allows the superposition of services, even at the mini-slot level by employing superposition coding at the transmitter and successive interference cancellation at the receivers \cite{song2017resource}, \cite{islam2016power}. Although most works on NOMA utilize the aspect of increased spectral efficiency to showcase superiority with respect to OMA, we further provide strong motivation for adopting NOMA as a conflict mitigation approach. An extensive set of numerical results, investigating NOMA's performance for both fixed and flexible numerology, shows the significant gains in terms of sum eMBB throughput, when adopting NOMA in a flexible numerology setting. 
\end{enumerate}

The rest of this paper is organized as follows: Section II presents the resource allocation optimization problem along with a novel formulation as a conflict minimization problem. Conflict-aware heuristic algorithms are proposed in Section III, including one inspired by a heuristic solution to the bin packing problem, while the problem re-formulation when using NOMA is presented in Section IV. Section V presents numerical results showing the near-optimal performance of the proposed heuristics as well as the superiority of NOMA for URRLC and eMBB coexistence, both in the case of flexible as well as fixed numerologies. Finally, conclusions are drawn in Section VI.

%% file: Sections/problem.tex
%We formulate the 2-D resource allocation problem.

\begin{table}
%label{notation}
\caption{Notation}
\renewcommand{\arraystretch}{1.4}
\setlength{\tabcolsep}{14pt}
\centering\label{ta:notation}
    $\begin{array}{|c|l|}
        \hline
        \textbf{Notation} & \textbf{Sets}  \\
        \hline
         \mathcal{K}^{(\ell)} & \text{Set of URLLC services} \\
         \mathcal{K}^{(c)} & \text{Set of eMBB services} \\
         \mathcal{K} & \text{Set of all services, for which } \mathcal{K} = \mathcal{K}^{(\ell)}{\cup}\mathcal{K}^{(c)} \\
         \mathcal{B} & \text{Set of candidate blocks with respect to the numerology} \\
         \mathcal{I} & \text{Set of all basic units of the grid, i.e., the minimum unit of}\\
            & \text{resource in the time-frequency domain (mini-slots)} \\
        \hline
        \textbf{Notation} & \textbf{Parameters} \\
        \hline
        \tau_{k} & \text{maximum delay tolerance of service } k \in \mathcal{K}  \\
        q_{k} & \text{throughput demand (in bits) of service } k \in \mathcal{K^{(\ell)}} \\
        \hline
    \end{array}$
\end{table}

We focus in this work on downlink scheduling, with one base station (BS) servicing both throughput hungry (eMMB) and ultra-low latency users (URLLC). The objective is to find the resource allocation in the time-frequency grid that maximizes the sum throughput of the former while satisfying the throughput demands and latency constraints of the latter. Our starting point is the system model of \cite{you2018resource}. We also utilized \cite{ch8e_x385_18} as a tool to implement the time-frequency grid.

The terminology employed in the rest of the paper is tabulated in Table \ref{ta:notation}:  $\mathcal{K}$ denotes the set of all services,  $\mathcal{K}^{(c)}$ the set of eMBB users, $ \mathcal{K}^{(\ell)} $ the set of URLLC users, $\mathcal{B} $ is the set of all possible resource blocks according to the numerology employed and finally,  $ \mathcal{I} $ denotes the set of all mini-slots.  We utilize the binary parameter $\alpha_{b,i}, b\in \mathcal{B}, i \in \mathcal{I}$ which indicates whether a block $b\in \mathcal{B}$ includes basic unit $i\in \mathcal{I}$, in which case $\alpha_{b,i}=1$, otherwise $\alpha_{b,i}=0$. Furthermore, we denote by $r_{b,k}, b\in \mathcal{B}, k \in \mathcal{K}$ the throughput of each resource block, under the constraint that the latency constraint is met; if the delay constraint is not met then the throughput is set to 0, i.e., in our throughput definition we incorporate the delay constraint. As such, the delay constraints of the URLLC services need not appear explicitely in the problem formulation presented in the following. %, i.e.,
%\begin{equation}
%r_{b,k}=\{\text{Capacity of block }b\text{ for service }k\} \times \mathbf{1}_{\{\tau_k-t_b>0\} }%\left\{
%\begin{array}{ll}
%0 & \text{if } \tau_k-et_b<0 \\
%\text{capacity} & \text{otherwise,}
%\end{array}
%\right.
%\end{equation}
%where $t_b$ is the end time of block $b$ and $\mathbf{1}_{\{x\}}$ is the indicator function for the logical proposition $x$. 
Additionally, by $x_{b,k}$ we denote a binary variable that takes the value $1$ if the resource block $b\in \mathcal{B}$ is assigned to service $k$, otherwise $x_{b,k}=0$. 

\begin{figure}[t]
\centering
\begin{subfigure}[b]{0.35\textwidth}
\centering
\includegraphics[width=\textwidth]{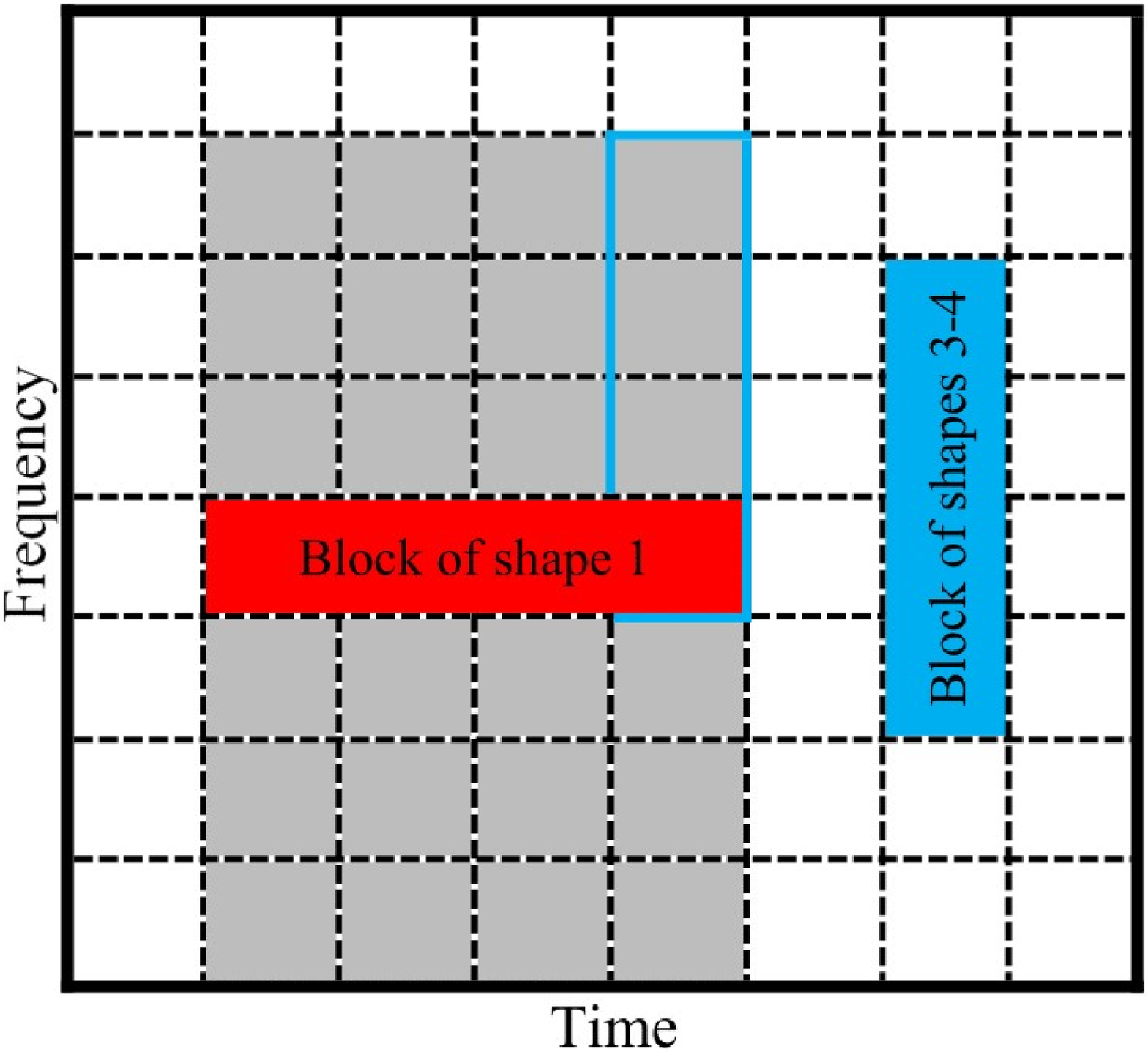}
\caption{Allocation of a candidate block of shapes 3-4, when a block of shape 1 already exists. Blue line indicates a conflicted block.}
\end{subfigure}
\begin{subfigure}[b]{0.35\textwidth}
\centering
\includegraphics[width=\textwidth]{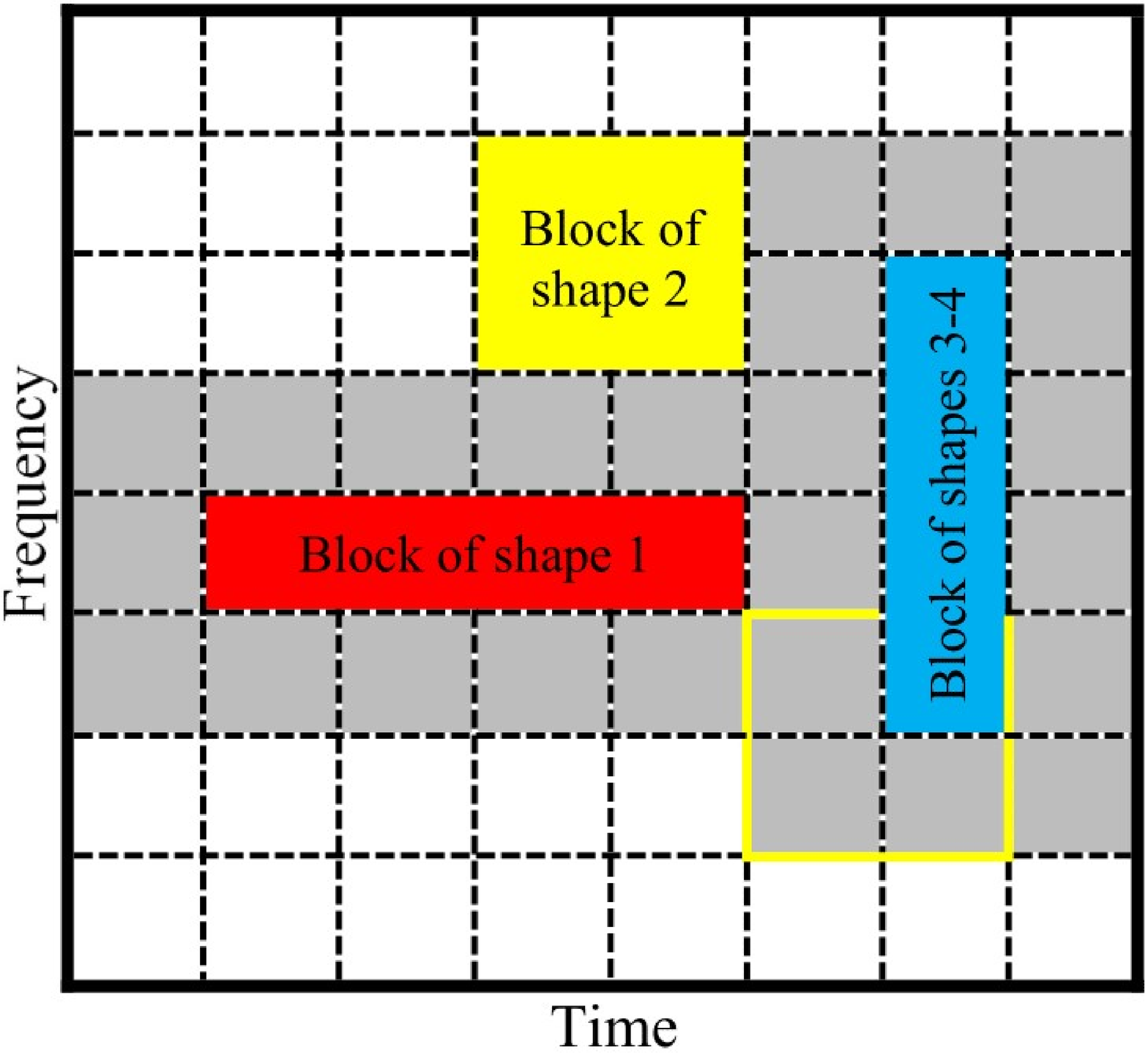}
\caption{Allocation of a candidate block of shape 2, when a block of shape 1 and a block of shapes 3-4 already exist. Yellow line indicates a conflicted block.}
\end{subfigure}
\caption{Time-frequency resource allocation, considering the flexible numerology context, with three types of resource blocks and the corresponding conflicts (grey).
}
\label{fig:block_alloc}
\end{figure}

In Table \ref{ta:flexible numerology} we describe the most widely utilized resource block specifications for 5G NR, depicted in Fig. 1(a) and (b); resource blocks of shape $1$ shown in red, resource blocks of shape $2$ shown in yellow and resource blocks of shapes $3-4$ shown in blue. Employing flexible numerology, $\mathcal{K}^{(c)}$ (eMBB) and $\mathcal{K}^{(\ell)}$ (URLLC) services have no restrictions and can utilize any of the given shapes. To demonstrate the concept of conflict, in Fig. \ref{fig:block_alloc}(a), we illustrate in gray shade the invalid placements for shapes 3-4 when a specific placement of shape 1 has taken place, while in Fig. 1(b) we show the invalid placements for blocks of shape 2, when an additional placement of shape 3-4 has been decided. 

\begin{table}[!t]
 \renewcommand{\arraystretch}{1.2}\centering \caption{Resource Blocks in Flexible Numerology}
\setlength{\tabcolsep}{6pt}
\centering \label{ta:flexible numerology}
\begin{tabular}{|l|c|c|c|c|}
\hline
\multicolumn{1}{|c|}{} &  \multicolumn{1}{c|}{Shape 1} & \multicolumn{1}{c|}{Shape 2} & \multicolumn{1}{c|}{Shape 3} & \multicolumn{1}{c|}{Shape 4}\\
\hline
TTI duration (ms) & 0.5 & 0.25 & 0.125 & 0.125\\
\hline
SCS (kHz) & 15 & 30 & 60 & 60\\
\hline
Symbol duration ($\mu$s) & 66.7 & 33.3 & 16.7 & 16.7\\
\hline
CP ($\mu$s) & 4.7 & 2.3 & 1.2 & 4.17\\
\hline
Number of Symbols & 7 & 7 & 7 & 6\\
\hline
\end{tabular}
\end{table}

A common objective in eMBB and URLLC coexistence is articulated in maximizing the sum throughput of $\mathcal{K}^{(c)}$ services under the constraint of satisfying the latency and throughput demands of $\mathcal{K}^{(\ell)}$, without any overlapping between the allocated resource blocks. In other words, our goal is to find the resource allocation that satisfies the URLLC users' demands, with minimal losses for eMBB users in terms of throughput, and, subsequently schedule all the remaining resource blocks to the eMBB services. The formal problem formulation is given as follows:
\begin{align}
\text{[P0]} \quad \max_{x_{b,k} \in \{0,1\}} &  \sum_{b\in\mathcal{B}}\sum_{k\in\mathcal{K}^{(c)}} r_{b,k}x_{b,k}, \label{eq:objective}\\
%& R_{total}-\sum_{b\in\mathcal{B}}\sum_{k\in\mathcal{K}^{(l)}} r_{b,k}x_{b,k}=\nonumber\\
%&R_{total}-\sum_{b\in\mathcal{B}}\sum_{k\in\mathcal{K}} cf_{b, k}r_{b,k}x_{b,k}\\
\text{s.t.} \quad & \sum_{b\in\mathcal{B}}r_{b,k}x_{b,k}\geq q_{k}, \quad k\in\mathcal{K}^{(\ell)} \label{eq:constraint1},\\
           & \sum_{b\in\mathcal{B}}\sum_{k\in\mathcal{K}}a_{b,i}x_{b,k}\leq 1, \quad i\in\mathcal{I} \label{eq:constraint2}.
\end{align}

%In \cite{you2018resource} authors proved that the problem is $\mathcal{NP}$-hard and provided a near-optimal solution. Despite the fact that the general target is to maximize the throughput of the $\mathcal{K}^{(c)}$ services, their proposed heuristic algorithm assigns the maximum throughput to the $\mathcal{K}^{(\ell)}$ services - until the (1b) constraint is satisfied - and after that assigns the remaining throughput to the $\mathcal{K}^{(c)}$ services. In our understanding, the existing methodology reveals the advantage of minimizing the blocks assigned to the $\mathcal{K}^{(\ell)}$ services, as it maximize the blocks assigned to the $\mathcal{K}^{(c)}$ services and also reduces the amount of the possible overlapping blocks between $\mathcal{K}^{c}$ and $\mathcal{K}^{(\ell)}$ services. Particularly it reveals that the maximization of the $\mathcal{K}^{(c)}$ services can be achieved by the maximization of the blocks assigned to the eMBB services. 

In \cite{you2018resource} it was proven that the combinatorial problem P0 is an ${NP}$-hard partition problem and a heuristic algorithm was proposed, referred to in the following as the \textit{baseline heuristic}. %; by employing Langrangian duality (LD) and linear programming (LP) relaxation, they proposed a re-formulation of P0, which (using optimization solvers) could attain near-optimal performance, at the cost of high complexity. 
%A heuristic algorithm was proposed by the authors of \cite{you2018resource}, 
The baseline heuristic uses a utility matrix $\mathbf{u}$ with elements $u_{b,k}$ that represent the utility of a block $b\in \mathcal{B}$ assigned to a specific service $k\in\mathcal{K}$. Then, in the \textit{first step} of the heuristic algorithm, the block $b$ is allocated to service $k\in \mathcal{K^{(\ell)}}$ with the maximum $u_{b,k}$ while all the overlapping -- to $b$ -- blocks are removed; notice that choosing the allocation that maximizes the utility without at the same time examining the ``cost'' of this placement in terms of generated conflict is clearly sub-optimal. The step is iterated until all the demands for $k \in \mathcal{K}^{(\ell)}$ are satisfied under the constraint (\ref{eq:constraint1}). Next, in the \textit{second step}, the placements for $k \in \mathcal{K}^{(c)}$ services are allocated, using a similar principle, %similarly to that of $\mathcal{K}^{(\ell)}$,
until no other non-overlapping blocks have remained. Hence, the placement of the $\mathcal{K}^{(\ell)}$ and $\mathcal{K}^{(c)}$ has been treated as two separate resource allocation problems. The complexity of the baseline heuristic algorithm was shown to be $\mathcal{O}(|\mathcal{B}||\mathcal{K}|\log(|\mathcal{B}||\mathcal{K}|))$, without accounting for the computation of utility matrices. 

The baseline heuristic has been extended in \cite{you2018resource} to incorporate other utility matrices denoted by $\mathbf{u}_{LP} \text{, }\mathbf{u}_{LD}\in{\mathbb{R}_{\mathcal{B}\times{\mathcal{K}}}}$, where $\mathbf{u}_{LP}$ and  $\mathbf{u}_{LD}$ denote the optimal solutions of the linear programming (LP) and the Lagrange dual (LD) relaxation of P0, respectively. With these two new utilities, an extension of the baseline heuristic was proposed to calculate concurrently the solution of the heuristic algorithm by adopting both $\mathbf{u}_{LP}$ and $\mathbf{u}_{LD}$ utilities and retaining the best result between them; this allowed to reach a near-optimal performance, at the cost of high computational complexity, especially considering that the dual problem P0-LD also applies a sub-gradient method. 

%Furthermore, using the utility values $u_{b,k}$, two further solutions were proposed. The first solution used the throughput $r_{b,k}$ of each pair $(b,k)$ as the utility. While, in the second,
%novel utility matrices $u_{LP} \text{, }u_{LD}\in{\mathbb{R}_{\mathcal{B}\times{\mathcal{K}}}}$ were introduced, 
%More precisely, ${u}_{LP}$ is the optimal solution of the LP problem, that derives from the relaxation of $x_{b,k}\in\{0,1\}$ to ${x_{b,k}}\in\left[0,1\right]$ in P0. Similarly, by relaxing the constraint (\ref{eq:constraint2}) that ensures no overlapping between blocks from the objective function, with the Lagrange multipliers, $\lambda_{i}\in{\mathcal{I}}$, the LD problem becomes, 

%\begin{align}
%    & \min_{\lambda\geqslant{0}}\Bigg\{ \max_{x_{b,k}\in\lbrace{0,1\rbrace}} \sum_{b\in{\mathcal{B}}}\sum_{k\in\mathcal{K}^{(c)}} r_{b,k}x_{b,k}+\sum_{i\in\mathcal{I}}\lambda_{i}\left(1-\sum_{b\in\mathcal{B}}\sum_{k\in\mathcal{K}}\alpha_{b,i}x_{b,k}\right)\Bigg\} \nonumber \\
%    & \text{s.t.} \quad \sum_{b\in\mathcal{B}}r_{b,k}x_{b,k}\geq q_{k}, \quad k\in\mathcal{K}^{(\ell)}. \nonumber %\\ 
%\end{align}
%The LD problem can be decomposed for the $\mathcal{K}^{(\ell)}$ and the $\mathcal{K}^{(c)}$ services and later to be solved with a subgrandient method \cite{palomar2006tutorial} to obtain the $u_{LD}$ matrix.

Discussing the above approach, whose basic principle (with few variations) can be found in other published work, e.g., \cite{anand2020joint}, we notice that despite the fact that the overall aim is to \textit{jointly} maximize the throughput of  $\mathcal{K}^{(c)}$ while meeting the demands of $\mathcal{K}^{(\ell)}$ services, these two interwoven goals are treated separately; in order to satisfy  constraint (\ref{eq:constraint1}), \textit{first} the demands of URLLC services are met and \textit{then} the placements of eMBB services take place.% Similar strategies were proposed in \textcolor{red}{CAN SOMEONE ADD A REFERENCE HERE?}. 

Such policies solve P0 by accounting only for constraint (\ref{eq:constraint1}), which is suboptimal as they do not consider the \textit{impact} of the $\mathcal{K}^{(\ell)}$ services allocation to the consequent allocation of the $\mathcal{K}^{(c)}$ services, i.e., constraint (\ref{eq:constraint2}). We notice that previously proposed algorithms operate on a single optimization target at any instance, that of maximizing first the URLLC throughout and then maximizing the eMBB throughput. Building on this observation, we will first show that the previously presented baseline heuristic  can be improved, if the conflict is taken explicitly into account.

%\begin{figure}
%    \centering
%    \includegraphics[width=0.5\textwidth]{Figures/aa1.png}   
%    \caption{Resource allocation of a candidate block and its corresponding conflicts; vertical blocks (light grey), horizontal blocks (grey) and square blocks (dark grey).}
%    \label{fig:block_alloc_conflicts}
%\end{figure}

To this end, we introduce an explicit description of the impact that the assignment of any resource block to a specific service has on the feasible assignments of the remaining blocks. In other words, we account for the amount of generated conflict by any specific URLLC or eMBB resource block placement. %To illustrate the idea, Fig. \ref{fig:block_alloc}a depicts all ``conflicts" that arise to an arbitrary placement of a resource block of shapes $3-4$, when a resource block of shape $1$ is already allocated in the grid; the specific block allocation \textit{forbids} any allocation for block $3-4$ in the sketched neighborhood in grey. Accordingly, \ref{fig:block_alloc}b depicts all ``conflicts" generating to a placement of a resource block of shape $2$, when a resource block of shape $1$ and a resource block of shapes $3-4$ are already allocated in the grid. These conflicts arise as constraint (\ref{eq:constraint2}) does not allow for overlapping (partial or full) of block placements. In light of this, it becomes evident that even if a particular resource block might have maximum (block) throughput, it's allocation could be sub-optimal due to the losses because of the generated forbidden placements around it, i.e., the generated conflicts described by constraint (\ref{eq:constraint2}) of P0 might outweigh the throughput gains.
To evaluate the impact of constraint (\ref{eq:constraint2}) explicitly, we define the conflict as
\begin{equation}\label{confl_par}
    c_{b,p}= \begin{cases}
    1, \text{  if } \sum_{b\in\mathcal{B}}\sum_{p\in\mathcal{B}}(\alpha_{b,i}+\alpha_{p,i})>1 \text{, } i\in\mathcal{I}, \text{ } b\neq{p}
    \\
    0, \hspace{12mm} \text{otherwise}
    \end{cases}
\end{equation}
for $b,p \in \mathcal{B}$.
As a next step we note that, 
%\begin{align}
%\sum_{b\in\mathcal{B}}\sum_{k\in\mathcal{K}^{(c)}} r_{b,k}x_{b,k}
%&=R_{total}-\sum_{b\in\mathcal{B}}\sum_{k\in\mathcal{K}^{(c)}} cf_{b, %k}r_{b,k}x_{b,k}\\
%\end{align}
\begin{align}
\sum_{b\in\mathcal{B}}\sum_{k\in\mathcal{K}^{(c)}} r_{b,k}x_{b,k}
&=R_{total}-\sum_{b\in\mathcal{B}}\sum_{p\in\mathcal{B}}\sum_{k\in\mathcal{K}} c_{b, p}x_{p,k}r_{b,k},
\end{align}
where $R_{total}$ denotes the maximum sum throughput of the whole resource grid with respect to $\mathcal{K}^{(c)}$ and the second triple sum represents the losses in $\mathcal{K}^{(c)}$ throughput, because of the conflicts generated by the placements of all services. %=\sum_{i\in \mathcal{I}}\sum_{b\in\mathcal{B}}\sum_{k\in\mathcal{K}} \alpha_{b,i}r_{b,k}$ is the overall throughput of the whole grid with respect to the eMBB users.
Given that $R_{total}$ is constant for any particular time-frequency grid realization,  the maximization of (\ref{eq:objective}) is equivalent to the minimization of the aggregate conflict, i.e.,

\begin{align}\label{,min_confl}
    %& \!\!\!\!\!\!\!\!\!\!\!\!\!\!\!\!\!\!\!\!\!\!\!\!\!\!\!\!\!\!\!\!\!\!\!\!\!\!\!\!\!\!\!\!\!\!\!\!\!\!\!\!\!\!\!\!\!\!\!\!\!\!\!\!\!\!\!\!\!\!\!\!\!\!\!\!\!\!\!\!\!\!\!\!\!\!\!\!\!\!\!\!\!\!\!\!\!\!
    %\!\!\!\!\!\!\!\!\!\!\!\!\!\!\!\!\!\!\!\!\!\!\!\!\!\!\!\!\!\!\!\!\!
    %\!\!\!\!\!\!\!
    %\max_{x_{b,k} \in \{0,1\}}  %\sum_{b\in\mathcal{B}}\sum_{k\in\mathcal{K}^{(c)}} r_{b,k}x_{b,k} %=\nonumber\\
    &\max_{x_{b,k} \in \{0,1\}} \left( R_{total}-\sum_{b\in\mathcal{B}}\sum_{p\in\mathcal{B}}\sum_{k\in\mathcal{K}} c_{b, p}x_{p,k}r_{b,k} \right) \Leftrightarrow\nonumber\\
    & 
    \min_{x_{b,k} \in \ \{0,1\}}\sum_{b\in\mathcal{B}}\sum_{p\in\mathcal{B}}\sum_{k\in\mathcal{K}} c_{b, p}x_{p,k}r_{b,k}.
\end{align}

\par Hence, the maximization of the sum eMBB throughput may be reduced to the minimization of the potential conflicts. We also note that:
\begin{align}
% \label{eq}
    \mathbb{E}\left[\sum_{b\in\mathcal{B}}\sum_{p\in\mathcal{B}}\sum_{k\in\mathcal{K}} c_{b, p}x_{p,k}r_{b,k}    \right]=\left|\mathcal{C} \right| \bar{r} ,\label{eq:average}
\end{align}
where $\mathbb{E}[\cdot]$ denotes expectation, $\mathcal{C}$ is the set of conflicts when all resource blocks have the same average throughput $\bar{r}=\mathbb{E}\left[ {r_{b,k}}\right] $ and $|\cdot|$ denotes cardinality;  i.e., from (6) and (7) it emerges that \textit{on a large grid we need, on average, to minimize the number of conflicts.} 

Considering these remarks, we propose novel heuristic algorithms  for P0, focusing on minimizing the number of placements of $\mathcal{K}^{(\ell)}$ services. The first set of heuristics, dubbed in the following as conflict-aware greedy, use ``conflict" enhanced variations of the utility proposed in the baseline heuristic and aim at closing the optimality gap. The second approach is built on an interpretation of (6) as a bin packing optimization problem \cite{Korte2006}; based on this approach we develop a lightweight scheduling approach that is shown to be near-optimal. %that is shown through numerical results to be near-optimal,
%with comparable performance to the computationally demanding LP-LD algorithm proposed in \cite{you2018resource}. 
%It is noteworthy that this approach also allows us to find feasible solutions at very low latency, e.g., at $0.5$ ms, for increasing URLLC demands.

Furthermore, as the minimization of conflicts is shown to be an equivalent optimization objective to the sum throughput maximization, we propose the use of NOMA to allow for overlapping of placements. The proposed heuristics and NOMA approaches are detailed in the next two sections.

%% file: Sections/algorithm.tex
% We demonstrated that the optimal scheduling to maximize the throughput of $\mathcal{K}^{(c)}$ services with moderate latency constraints while accommodating throughput demands of $\mathcal{K}^{(\ell)}$ services, is equivalent to minimizing the number of $\mathcal{K}^{(\ell)}$ placements and also minimizing the total throughput loss of the former.

\subsection{Conflict-aware heuristic solutions}
We first propose extensions of the baseline heuristic, in \cite{you2018resource}, by introducing penalties in URLLC resource allocations, expressed as functions of the conflict. %The penalties describes the resulting conflicts impact of URLLC services placement on eMBB allocation efficiency.
To this end, we introduce two metrics for the conflict induced by $\mathcal{K}^{(\ell)}$ services allocation. The aggregate conflict $C^{t}_{b}$,  
\begin{equation}\label{confl_1}
    C^{t}_{b} = \sum_{p\in\mathcal{B}}c_{b,p}, \text{ } p,b\in\mathcal{B}, 
\end{equation}    
that measures the total number of overlapping blocks with the block $b$, and, the average conflict $C^{r}_{b,k}$,
\begin{equation}\label{confl_2}
    C^{r}_{b,k} = {\sum_{p\in\mathcal{B}}}\frac{c_{b,p}r_{p,k}}{C^{t}_{b}}, \text{ }p,b\in\mathcal{B} \text{ and } k\in\mathcal{K}^{(\ell)}
\end{equation}
that corresponds to the average throughput -- for every service $k\in\mathcal{K}^{(\ell)}$ -- of the blocks $p\in\mathcal{B}$ that overlap with block $b\in\mathcal{B}$.
%\footnote{We note that despite that the average throughput - of all services $k\in\mathcal{K}$ - of the blocks $p\in\mathcal{B}$ that overlap with block $b\in\mathcal{B}$, $C^{r}_{b}=\sum_{k\in\mathcal{K}}{\sum_{p\in\mathcal{B}}}\frac{c_{b,p}r_{p,k}}{C^{t}_{b}},$ should seem a more ``global" conflict measure. We do not use this metric, as it is much more computational demanding, especially for a large amount of services.} 

Using these new conflict measures, we propose three variations for the utility matrix to be used in solving P0:
\begin{itemize}
\item In the first version the utility becomes, $$u^{total}_{b,k}=\frac{r_{b,k}}{C_{b}^{t}};$$
\item In the second variation, the utility becomes, $$u^{avg}_{b,k}=\frac{r_{b,k}}{C_{b,k}^{r}};$$
\item Finally, in the third variation, we use the following utility,
$$u^{last \text{ } pl.}_{b,k}= \begin{cases}
r_{b,k}, \hspace{12mm} \text{if } k=1,\dots,\left|\mathcal{K}^{(\ell)}\right|-1 
\\
u^{avg}_{b,k}, \hspace{12mm} \text{if }  k=\lbrace{\left|\mathcal{K^{(\ell)}}\right|\rbrace}.
\end{cases}$$
\end{itemize}
The utility matrix $u^{last \text{ }pl.}_{b,k}$ is introduced to incorporate a ``compromise'' between the baseline and the conflict-aware approaches; notably, it considers the impact of the conflict only in the last $\mathcal{K}^{(\ell)}$ service placement, since our simulations revealed that in this last placement, usually, more blocks are required to satisfy the demands constraint.

\begin{algorithm}[t]
\caption{Conflict-aware Resource Allocation Algorithm (\textit{CA}) based on \cite{you2018resource}}
\label{alg0}
\begin{algorithmic}
\REQUIRE 
$\mathbf{u}^{(\ell)}=[u_{b,k}], \text{ }b\in \mathcal{B}, k\in \mathcal{K^{(\ell)}}$, utility matrix for $K^{(\ell)}$  ($\mathbf{u}^{total}$, $\mathbf{u}^{avg} \text{ or } \mathbf{u}^{last \text{ } pl.}$) and 
$\mathbf{u}^{(c)}=[r_{b,k}], \text{ }b\in \mathcal{B}, k\in \mathcal{K}^{(c)}$, utility matrix for $K^{(c)}$.
\ENSURE Block-service assignment $\mathbf{s}$.

\REPEAT 
%\qquad \qquad $\blacktriangleright$ Resource allocation of the $\mathcal{K}^{(\ell)}$ services 
\STATE Remove from $\mathcal{B}$ the blocks in $\mathbf{s}$ and the overlapping with $\mathbf{s}$ blocks.
\STATE $(b^{'},k^{'})\leftarrow\argmax_{b\in\mathcal{B},{k}\in{\mathcal{K}^{(\ell)}}} u_{b,k}^{(\ell)}, \quad \mathbf{s}\leftarrow\mathbf{s}\cup\lbrace{(b^{'},k^{'})}\rbrace.$ 
\IF {$q_{k^{'}}$ \text{ is met } }  
\STATE {$\mathcal{K}^{(\ell)}\leftarrow{\mathcal{K^{(\ell)}}\setminus{k^{'}}}.$}
\ENDIF
\UNTIL{$\mathcal{K^{(\ell)}}=\varnothing$ \textbf{or} $\mathcal{B} = \varnothing$} 
\IF{$\mathcal{K}^{(\ell)}\neq\varnothing$}
\STATE{The demand of the remaining users in $\mathcal{K}^{(\ell)}$ can not be met.}
\ENDIF
\REPEAT
%\qquad \qquad \qquad \qquad \qquad $\blacktriangleright$ Resource allocation of the $\mathcal{K}^{(c)}$ services 
\STATE{Remove from $\mathcal{B}$ the blocks in $\mathbf{s}$ and the overlapping with $\mathbf{s}$ blocks.}
\STATE{$(b^{'},k^{'})\leftarrow\argmax_{b\in\mathcal{B},{k}\in{\mathcal{K}^{(c)}}} u_{b,k}^{(c)}, \quad \mathbf{s}\leftarrow\mathbf{s}\cup\lbrace{(b^{'},k^{'})}\rbrace.$}
\UNTIL{$\mathcal{B}=\varnothing$}
\end{algorithmic}
\end{algorithm}

Concluding, the set of conflict-aware heuristics allocate the URLLC services by using the $\mathbf{u}^{total}$ or $\mathbf{u}^{avg}$ or $\mathbf{u}^{last \text{ }pl.}$ utility matrices, while for the eMBB services the utility remains unchanged. The heuristics are outlined in Algorithm \ref{alg0} as pseudo-code.

% Furthermore authors in \cite{you2018resource} denote a utility matrix $u_{b,k}$. The utilities $u_{LP}$ and $u_{LD}$ are the optimal solutions of the Linear Programming (LP) relaxation and the Lagrange Dual (LD), respectively. Later, an alternative solution to the baseline algorithm is to allocate each block to the service with the maximum utility, either using $u_{LP}$ or $u_{LD}$. Then the best alternative is the solution "LP+LD" that gives the best results between $u_{LP}$ and $u_{LD}$. Accordingly, we integrate the former "LP+LD" methodology to our heuristic approach; that incorporate the conficts.

\subsection{Heuristic inspired from bin packing optimization}
In  the  standard  bin  packing  problem  formulation, the goal is to find the optimal placement of items of different volumes in the minimum number of containers (bins) of fixed volume \cite{Korte2006}. Although the bin  packing is a combinatorial $NP$-hard problem, due to it’s  widespread  encounter in a large number of settings, various proposed heuristics have been reported in the literature  with different optimality gaps. Here, we propose a novel, computationally efficient scheduling approach, inspired by the refined-first-fit heuristic for the standard bin packing problem.

The proposed scheduling heuristic that accounts for conflicts is summarized in Algorithm \ref{alg1}, %performing allocation of blocks to $\mathcal{K}^{(\ell)}$ services by 
jointly minimizing the number of $\mathcal{K}^{(\ell)}$ resource allocations (placements) and throughput losses for $\mathcal{K}^{(c)}$ users. %Algorithm \ref{alg1} finds a near-optimal allocation of resource blocks  that satisfies all the demands of $\mathcal{K}^{(\ell)}$ while accounting for the  $\mathcal{K}^{(c)}$ services' throughput losses. 
% the throughput loss effects caused by the overlapping blocks. 
%In this way more resource blocks remain for resource allocation to $\mathcal{K}^{(c)}$ services. 
Allocation of resources to $\mathcal{K}^{(\ell)}$ services and $\mathcal{K}^{(c)}$ services is treated sequentially but still in an interwoven approach, with URLLC being served first to meet the latency requirements.
 %: Matrix $r$ of size $|B| \times | \mathcal{K}^{(\ell)} |$ denotes the throughput matrix for all pairs of blocks and services. An element $r(b, k)$ represents the achievable throughput of a block-service pair $(b, k)$ $(b \in B$ and $k \in K)$. 
In the following, the vector $\mathbf{e}$   of length $|\mathcal{B}|$ has as elements the aggregated throughput losses for each allocation of a block  $b \in \mathcal{B}$, i.e.,
\begin{align}
\label{eq}
e_b=\sum_{p\in\mathcal{B}}\sum_{k\in\mathcal{K}^{(c)}} c_{b, p}r_{b,k}.
\end{align}

\begin{algorithm}[t]
\caption{Bin Packing Based Resource Allocation Algorithm (BPB)}
\label{alg1}
\begin{algorithmic}
\REQUIRE throughput matrix $\mathbf{r}=[r_{b,k}], \quad b\in \mathcal{B}, k\in \mathcal{K}$, aggregated-throughput-loss vector $\mathbf{e}$, demand vector of URLLC services $\mathbf{q}$, set of all available resource blocks $\mathcal{B}$, overall number of categories (bins) $H$.
\ENSURE Block-service assignment $\mathbf{s}$.

\FOR{$ k=1$ to $|\mathbf{q}|$} 
\STATE create the following categories:
\FOR{$ i=1$ to $H$} 
% \IF{i=10}
% \STATE $Cat^i U^k=$ all resource blocks $b \in B$ where \\ $\lceil{q^k/r(b, k)}\rceil >= 10$;
% \ELSE
\STATE $Cat^i U^k=$ all resource blocks $b \in \mathcal{B}$ where \\ $\lceil{q_{k}/r_{b,k}}\rceil = i$;
% \ENDIF
\STATE Check pairwise conflicts among categorized blocks and remove the blocks with the higher aggregated-throughput-loss;
\ENDFOR
\ENDFOR
\\
$\blacktriangleright${\textbf{\quad Phase ($\mathcal{K}^{(\ell)}$ resource allocation):}}
% \REPEAT
% \WHILE{<condition>} 
\FOR{$i=1$ to $H$} 
\STATE select the $Cat^i U^k$ which has the least number of blocks;
\IF{ ($|Cat^i U^k| \geq i$ and $q_{k}$ is not already met)}
\STATE $ \mathcal{B} ^ \prime \leftarrow $ (select $i$ number of blocks in $Cat^i U^k$ with the least aggregated-loss-value);
\STATE	$\mathbf{s} \leftarrow \mathbf{s} \cup {(b ^ \prime, k ^ \prime)}$ , $k ^ \prime=i , \forall b ^ \prime \in \mathcal{B} ^ \prime$;
\STATE Remove from $\mathcal{B}$ the blocks in $\mathbf{s}$ and those overlapping with the blocks in $\mathbf{s}$;

\IF{ $q_{k}$ is met}
\STATE $\mathcal{K}^{(\ell)} \leftarrow \mathcal{K}^{(\ell)} \backslash \{k ^ \prime \}$;
\ENDIF
\ENDIF
\ENDFOR
% \UNTIL{$\mathcal{K}^{(\ell)}=\emptyset$ or $B=\emptyset$}
% \ENDWHILE
\\
$\blacktriangleright${\textbf{\quad Phase ($\mathcal{K}^{(c)}$ resource allocation):}}
\REPEAT 
\STATE $(b ^ \prime, k ^ \prime) \leftarrow \arg\max_{ b \in \mathcal{B}, k \in \mathcal{K}^{(c)}}  r_{b,k}$;
\STATE  $\mathbf{s} \leftarrow \mathbf{s} \cup {(b ^ \prime, k ^ \prime)}$;
\STATE Remove from $\mathcal{B}$ the blocks in $\mathbf{s}$ and those overlapping with the blocks in $\mathbf{s}$;
\UNTIL{ $\mathcal{B} = \emptyset$ }
\end{algorithmic}
\end{algorithm}

% \textcolor{blue}{all $\mathcal{K^{(\ell)}}$ have the same demand q}
% \textcolor{red}{\textit{Is this what we are doing? I can't remember:} In Algorithm \ref{alg1}, first, all the services $k \in \mathcal{K^{(\ell)}}$ are ordered in decreasing demand size $q_k$. Starting with the URLLC service with the highest demand $q_k, k \in \mathcal{K^{(\ell)}}$,}  

The proposed heuristic works as follow: for each $k \in \mathcal{K^{(\ell)}}$ we generate $H$ categories (bins) with decreasing fractional sizes with respect to $q_k, k \in \mathcal{K^{(\ell)}}$, i.e., category $i\in \{1,\ldots,H\}$ is defined as the set of all
% the throughput values of  
resource blocks $b \in \mathcal{B}$ for which the ceiling of the service demand ratio over the throughput of block $b$ is equal to $i$, or equivalently, category $Cat^i U^k$ contains the available resource blocks which satisfy at least $1/i$-th of the service demand $q_k$. Formally, we define% for $k$, $ k =1,\ldots, | \mathcal{K}^{(\ell)} |\} $, $ i \in \{1, \ldots, M\} $, 

\begin{align}
Cat^i U^k&=\left\{b: \Biggl\lceil{\frac{q_{k}}{r_{b,k}}}\Biggr\rceil = i, \forall b \in \mathcal{B}\setminus\{Cat^{j}U^k\}_{j=1,\ldots,i-1}\right\}, \nonumber\\
& k\in \mathcal{K^{(\ell)}}, i\in \{1,\ldots, H\},
\end{align}
where $\lceil x\rceil$ denotes the smallest integer bigger or equal to $x$.

\begin{figure*}[t]\label{fixed_flexible}
    \centering
            \includegraphics[width=1\textwidth]{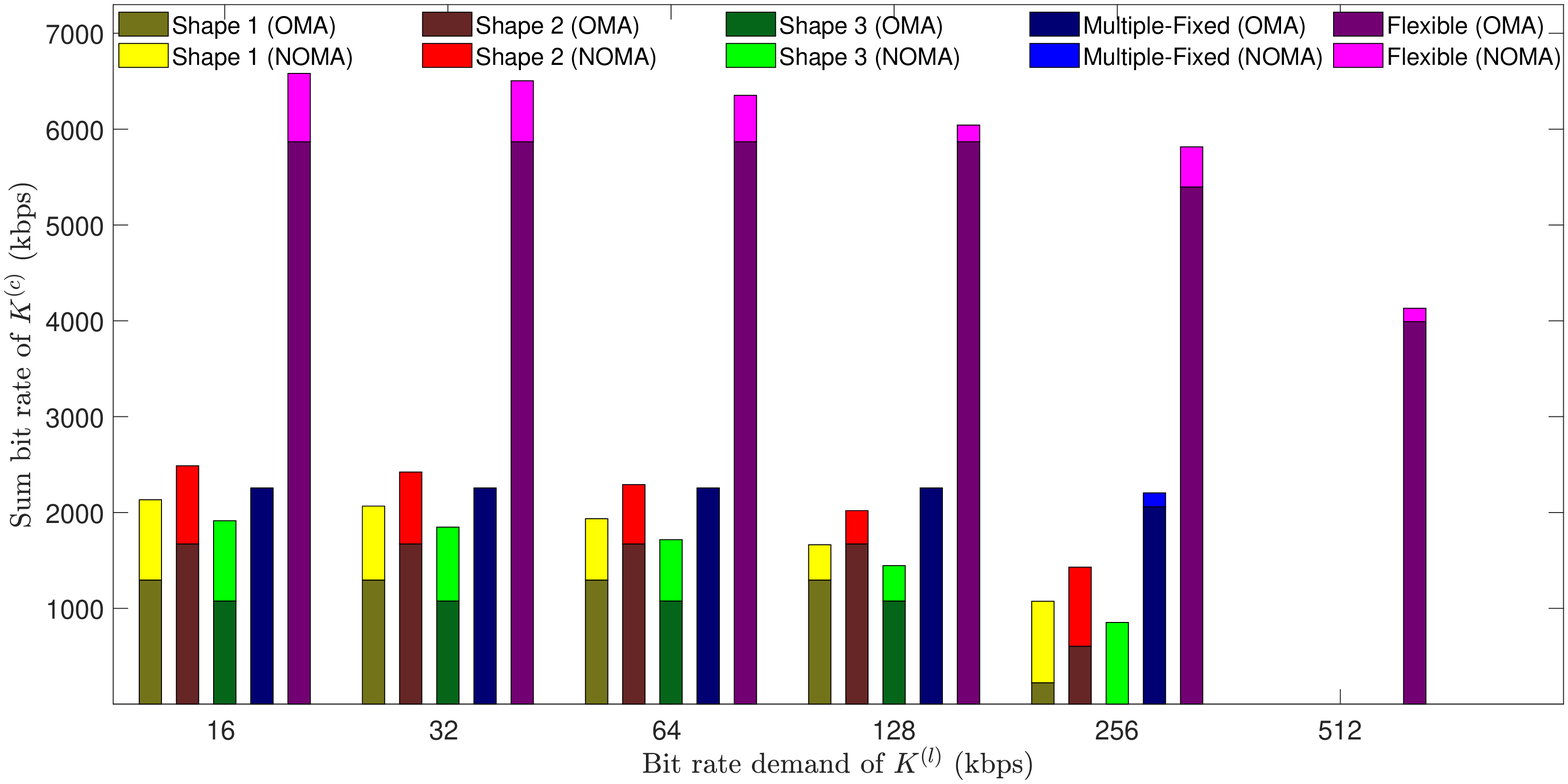}
    \caption{Sum bit rate for $\mathcal{K}^{(c)}$ services when employing NOMA and OMA for fixed, multiple and flexible numerology, under several $q_{k}$ data demands and delay tolerance value $\tau_{k}=1$ msec, $k\in\mathcal{K}^{(\ell)}$. The lighter colors depict the NOMA sum bit rate gains in comparison to the OMA. Fixed and multiple-fixed numerologies result in infeasible outputs for $q_{k}=256$ kbps and $q_{k}=512$ kbps, i.e., it is infeasible to satisfy all URLLC demands using these numerologies. On the other hand, flexible  numerology does not suffer from infeasibility even for $q_{k}=512$ kbps. The tremendous gains in using flexible numerology are consistent across all service demand scenarios. The gains in using NOMA are more accentuated in lower URLLC demands.}
    \label{fig:fixed_flexible}
    \end{figure*}

For example, $Cat^1 U^1$ is the category of the blocks which individually satisfy the whole demand of the URLLC service $k=1$. Therefore, the categories created for service $k \in \mathcal{K^{(\ell)}}$ range from $Cat^1 U^k$ -- containing the most valuable blocks (valuable in terms of throughput $r_{b,k}$) -- till $Cat^{H} U^k$, containing the least valuable blocks in order. Note that i) we need \textit{at most} $i$ elements from $Cat^iU^k$ to satisfy the demand $q_{k}$ of service $k \in \mathcal{K^{(\ell)}}$; 
% ii) once a resource block is allocated into a category $\#i$, it cannot be placed in a category of higher order $\#j$ with $j>i$, i.e., each resource block is assigned to one category; (a resource block which satisfies $1/i$-th of a service k may be able to satisfies $1/j$-th of another service; therefore a resource block might be in more than one category. to avoid the over utilization of a block, we aliminate it from all of its containing categories once it is allocated to a service in next phase of the alg)
ii) categories might be empty, so $H$ needs to be defined according to the expected throughput per mini-slot as well as its variance. %For example, in our numerical results provided in Section V, we have set $M=10$.  

Inside each category, we subsequently introduce a further minimization problem in order to select the elements from each category that incur the minimum loss to eMBB, i.e.,
\begin{align}
    \min_{y_b\in\{0,1\}} & \sum_b {e_b}{y_b}, \text{ } b \in (\mathcal{K}^{(l)} \cap {Cat^iU^k})\\
    %\text{s.t.} \quad & \sum_b{\mathbf{1}_{\{ b \in \mathcal{K}^{(l)} \cap {Cat^iU^k}\}}} \leq i, \quad b\in {Cat^iU^k} \nonumber.
    \text{s.t.}& \sum_{b\in{{Cat^iU^k}}}y_b \leq i \nonumber. 
\end{align}
%\textcolor{red}{Ersi: does anyone have any idea of how to write (13) ? I want to minimize the sum loss but taking only i elements,Sotiris: i tried to write it in the form of a knapsack, i think it is ok, but not absolutely sure.}
Note that if (12) is interpreted as a knapsack problem, each element of a given category has the same weight (equal to unity), while the values (losses in the specific instance) differ. 
Similar problems are encountered in different settings, e.g., the subcarrier resource allocation in \cite{MiroEurasip}. Exploiting these previous results, we reproduce a simple heuristic according to which the elements of each category are \textit{re-ordered}\footnote{The ordering has a complexity $\mathcal{O}\big(\max_{i, k}\{|Cat^iU^k|\log(|Cat^iU^k|)\}\big)$.} in increasing aggregated loss $e_b, b \in \mathcal{B}$. Subsequently, the first $i$ elements of category $Cat^iU^k$ are allocated to URLLC.

As an example, after this step, the first element of $Cat^1U^k$ is the resource block that can simultaneously cover the demand $q_{k}$ of URLLC service $k$ while incurring the least aggregate losses for the eMBB users.  The joint minimization of the number of $\mathcal{K}^{(\ell)}$ placements and the losses due to conflicts is achieved simply by assigning to service $k \in \mathcal{K^{(\ell)}}$ the first $i$ elements of $Cat^iU^k$, starting from $i=1$, i.e., the allocation for demand $q_{k}$ starts from $Cat^1 U^k$. As explained before, the most valuable  categories in terms of throughput satisfy URLLC services by using the least number of resource blocks and result in the minimum number of $\mathcal{K}^{(\ell)}$ placements, that is expected on average to incur the minimum losses due to conflicts. Furthermore, having re-ordered the elements of each category in increasing eMBB loss value, we jointly account for both constraints (\ref{eq:constraint1}) and (\ref{eq:constraint2}) in one go.
%It iteratively selects the category with the least number of blocks and defines the specific service $k^ \prime=k$, indicated by the selected category, to be resource allocated. Then, in selected category $cat^i U^k$, $i$ number of blocks which have the least aggregated-loss-value are selected to be allocated to  $k^ \prime$. 
After each allocation, the allocated blocks are removed from $\mathcal{B}$ and all other categories. This procedure is repeated until the demand of all of the $\mathcal{K}^{(\ell)}$ services are satisfied or no more blocks remain in the categories.

In the last phase of the algorithm, the resource allocation to $\mathcal{K}^{(c)}$ services takes place. This is performed by selecting the block-service pairs  with the highest throughput $r_{b, k}, b\in \mathcal{B}, k\in \mathcal{K}^{(c)})$ from the \textit{remaining} available blocks. The latter have not been allocated to a URLLC service, since once a block is allocated it is removed from $\mathcal{B}$. This step is iterated until no more blocks remain available. % or no more $\mathcal{K}^{(c)}$ services remain unallocated.

Finally, we consider a modified version of the bin packing based heuristic (mBP), targeting on challenging time-frequency grids, where infeasibility is the major issue. In this case, we introduce a pre-processing step to check the feasibility of the grid. We first count the total throughput of all available block placements and compare with throughput resulting from the placement of all the available blocks for the URLLC services, in both cases with respect to the constraint (3). Then if,  
$$\sum_{b'\in\mathcal{B}}\sum_{k\in\mathcal{K}^{(\ell)}} r_{b',k}>\delta\sum_{b'\in\mathcal{B}}\sum_{k\in\mathcal{K}} r_{b',k},$$
where $b'\in\mathcal{B}$ are the blocks that satisfy constraint (3) and $\delta\in(0,1)$, instead of using the $e_b$ metric for the allocation of the $k\in\mathcal{K}^{(\ell)}$ services we switch the metric to $e'_{b}=\max{r_{b,k}}$, $k\in\mathcal{K}^{(\ell)}$, in order to ensure the URLLC's services allocation.

%% file: Sections/results.tex
In this section, we present numerical results for both OMA and NOMA schemes, for different 5G URLLC configurations and numerologies; fixed, multiple-fixed and flexible numerology. This exercise allows us to highlight the importance of flexible numerology, while motivating NOMA as a conflict mitigation approach. Here, we mainly focus on the conflicts aspect, rather than on deployment, feasibility or coordination issues, which are important enough to deserve an independent study. We then move on to a comparative analysis of the proposed heuristic Algorithms \ref{alg0} (conflict aware, CA) and \ref{alg1} (bin packing based, BPB) for OMA, with a goal to support the potential of the proposed conflict aware scheduling.    

We use the simulation setup given in \cite{you2018resource}, implemented based on the control channel overhead model for supporting the flexible numerology defined in \cite{miao2017physical} and considers the effect of guard band (i.e., of the cyclic prefix) on the achievable data rate by blocks, as modeled in \cite{yazar2018flexibility}. The computation of the achieved throughput per block $r_{b,k}$ relies on the configuration of block $b$ (see Table II), with a total number of nine multipath channel profiles \cite{channel}, calculating the throughput based on the model introduced in \cite{channel_2}; for URLLC users the throughput values incorporate the delay constraints so that non-zero throughput is available only in these block in which the delay constraint is met. The throughput model also considers intersymbol-interference (ISI) depending on CP, and approximates the inter-channel interference (ICI) between the neighboring subbands of different numerologies.        
\par In detail, regarding the simulation parameters, we assume a time-frequency grid with a 2 msec and 2 MHz domain (i.e., of dimesnions $16\times11$). As a result, we have a set of $\mathcal{I}=\{1,\dots,176\}$ mini-slots and a corresponding set of $\mathcal{B}=\{1,\dots,549\}$ candidate blocks with respect to the numerology, where every candidate block consists of 4 elements of $\mathcal{I}$. The resource  block details are given in Table II. Blocks of shape 1 ($4\times1$), $\mathcal{B}_1\subset{\mathcal{B}}$, include a multitude of $|\mathcal{B}_1|=143$ resource blocks. Blocks of shape 2 ($2\times2$), $\mathcal{B}_2\subset{\mathcal{B}}$, include a multitude of $|\mathcal{B}_1|=150$ resource blocks. Finally, blocks of shape 3 and 4 ($1\times4$), $\mathcal{B}_3,\mathcal{B}_4\subset{\mathcal{B}}$ include the same multitude of blocks $|\mathcal{B}_3|=|\mathcal{B}_4|=128$. Furthermore, we consider 10 users in total, 5 URLLC and 5 eMBB, with $|\mathcal{K}^{(c)}| = |\mathcal{K}^{(\ell)}| = 5$. Moreover, the chosen latency tolerance and bit rate demands for the URRLC users are $\mathbf{\tau}=\{0.5, 1, 1.5, 2, 3\}$ msec and $q=\{16,32,64,128,256,512\}$ kbits/sec (kbps), respectively. The latency tolerance for the eMBB users is fixed and equal to $\mathbf{\tau}=2$ msec. The SNR range is generated by numbers uniformly distributed in the interval $[5,30]$ dB. Our references to the ``optimal solution'' in the following text correspond to the solutions provided by the Gurobi optimization solver and are used as a benchmark for the optimality gap of the proposed heuristics. Finally, the outputs of all the simulation results are assessed over $N=1000$ Monte Carlo simulations. 

\subsection{Performance comparison between NOMA and OMA scheduling under different numerologies}

First, we compare the outcome of OMA and NOMA schemes for different numerologies. In the case of fixed numerologies, shape 1 (horizontal), shape 2 (square) and shape 3 (vertical) type of blocks are considered separately. Furthermore, capturing a common scenario in practical systems, we define as the multiple-fixed numerology the one in which eMBB uses resource blocks of shape 1 (horizontal) and URLLC of shape 3 (vertical). Finally, in the case of flexible numerology all type of shapes, given in Table II, are available to all services. % From these results it becomes apparent that flexible numerology in combination with NOMA can offer distinct gains across varying URLLC demands. Notably, as the URLLC demands increase flexible numerology is the only approach that avoids infeasibility issues, i.e., not covering all of URLLC demnads. 

In Fig. \ref{fig:fixed_flexible}, the sum bit rate for the eMBB services, $\mathcal{K}^{(c)}$ when applying the optimal i) NOMA and ii) OMA scheduling are shown. The  NOMA sum bit rate gains to the OMA are depicted with the lighter color in each bar. The latency tolerance and bit rate demands considered are $\tau=1$ msec and ${q}=\{16, 32, 64, 128, 256, 512\}$ kbps, respectively, for five $\mathcal{K}^{(\ell)}$ and five $\mathcal{K}^{(c)}$ users. In all cases, as expected, flexible numerology significantly outperforms the fixed and multiple-fixed numerology. Moreover, multiple-fixed overpasses the performance of fixed numerology in the OMA case. From these results it becomes apparent that flexible numerology in combination with NOMA can offer distinct gains across varying URLLC demands. Notably, as the URLLC demands increase, flexible numerology is the only approach that avoids infeasibility issues, i.e., not covering all of URLLC demands. 

Focusing on the comparison between OMA and NOMA, the NOMA consistently outperforms OMA. More precisely, NOMA based scheduling is shown to increase particularly the sum throughput of eMBB users under fixed numerology, although NOMA also improves the overall performance when using flexible numerology as well. On the other hand, NOMA does not affect the performance under multiple-fixed numerology; this is due to the fact that in the specific grid used in the simulations, overlapping of blocks is limited in the case of multi-fixed numerology.

%flexible and multiple-fixed numerology significantly outperform the fixed structure, and NOMA outperforms the optimal OMA scheduling. The gains are more accentuated at low delay tolerance values, which indicates that NOMA can be beneficial for ultra-low-latency, a scenario of significant practical importance, e.g., in industry 4.0 or vehicle to everything (V2X) communications. The amount of improvement Flexible structure provides a reduced amount of improvement over the multiple-fixed  structure.   

Furthermore, in Fig. \ref{fig:gaps}, the normalized to NOMA performance gap between OMA and NOMA (expressed as a percentage) is shown, for different numerologies. The superiority of NOMA is reconfirmed both for fixed and flexible numerology, for different values of the URLLC latency tolerance $\tau_{k} = \{0.5, 1, 1.5, 2\}$ msec, $k\in\mathcal{K}^{(\ell)}$. Finally, in the case of flexible numerology, the lower the delay tolerance $\tau_k$, the higher the gains in using NOMA as opposed to OMA. 
%The above outcome, is strongly related with the reduction of the available block placements in the case of fixed numerology, which become even more restrictive under the latency tolerance constraint. 
The performance fluctuations, illustrated in Fig. \ref{fig:gaps}, are strongly related to the different values of the bit rate demands $q_k, \text{ }k\in\mathcal{K}^{(\ell)}$. More precisely, after a close inspection of the simulation outputs, we came to the conclusion that the gap between the demand of a service $k\in\mathcal{K}^{(\ell)}$ and the achievable throughput of the block, in which the service is allocated, plays an important role. A higher gap between the two corresponds to a decisive reduction of the overall available throughput for the scheduling of the $\mathcal{K}^{(c)}$ services in the OMA case, which in turn offers a crucial advantage to the NOMA scheme that allows overlaps. 
 
In Fig. \ref{fig:pyt} the overall scheduling output of all services is depicted in the case of OMA and NOMA, for $\tau_{k}=\{0.5, 1, 2\}$ msec and $q_{k}=\{16, 32, 64, 128, 256, 512\}$ kbps, respectively, for all $k\in\mathcal{K}^{(\ell)}$. In the case of OMA and a small value of $q_k, \text{ }k\in \mathcal{K}^{(\ell)}$ depicted in Fig. \ref{fig:pyt}(a), overlapping of resource blocks is not allowed, while on the other hand in the case of NOMA, depicted in Fig. \ref{fig:pyt}(b), the opportunity of overlapping resource blocks increases the available resource blocks for eMBB allocation; notice also that the choice of resource blocks assigned to URLLC is different. Similar outcomes are depicted in Figs. \ref{fig:pyt}(c), (e), OMA case, and Figs. \ref{fig:pyt}(d), (f), NOMA case, in which higher values of $q_k, \text{ }k\in\mathcal{K}^{(\ell)}$ are considered. In this case, though, to meet the higher bit rate demands of the URLLC services, more resource blocks are allocated to them, e.g., for $q_k=512$ kbps (in Figs. \ref{fig:pyt}(e) and (f) almost half of the resource blocks are used to cover URLLC demands.

%
%In this Section, we present results both for the heuristic algorithms in the case of OMA as well as in NOMA. The performance of the proposed heuristic Algorithms \ref{alg0} and \ref{alg1} are evaluated for different 5G URLLC configurations and numerologies.
%To showcase the effectiveness of the proposed heuristic resource allocation algorithms, we compare their performance against the global optimum (estimated using Gurobi optimization solvers) and the baseline LP-LD algorithm discussed in Section II, using the same simulation setup as in \cite{you2018resource}\footnote{We thank the authors of \cite{you2018resource} for kindly sharing their simulation codes in IEEE DataPort.}.

%This simulation environment was implemented based on  the control channel overhead model for supporting the flexible numerology defined in \cite{miao2017physical} and considers the effect of guard band (i.e., of the cyclic prefix) on the achievable data rate by blocks as modeled in \cite{yazar2018flexibility}. We measured the bit rates per user in $\mathcal{K}^{(c)}$ for URLLC latency tolerance values $\mathbf{\tau}=\{0.25, 0.5, 1, 1.5, 2\}$ ms and bit rate demands $\mathbf{q}=\{16, 32, 64, 128, 256, 512, 1024\}$ kbps for \textcolor{black}{five} services $k \in \mathcal{K}^{(\ell)}$.

\subsection{Performance of proposed heuristic algorithms}
Although NOMA clearly outperforms OMA, its use might be prohibited by a number of factors, including the need for multiple decoding steps and the impact of imperfect SIC. As a result, the evaluation of OMA scheduling approaches is paramount. 
In this subsection we discuss the proposed heuristics. As a validation step, we first evaluate and compare the optimality gaps of the baseline heuristic (presented in \cite{you2018resource}) and the proposed conflict aware heuristics with utilities ($\mathbf{u}^{total}$, $\mathbf{u}^{avg}$ and $\mathbf{u}^{last \text{ }pl.}$), denoted by $CA(\cdot)$ with input one of the corresponding utility matrices, against the global optimum of P0. Then, we provide additional results with all proposed heuristics employing flexible numerology. 

%\begin{figure}[h]
%\centering
%\begin{subfigure}[b]{\linewidth}
%    \centering
%    \includegraphics[width=0.50\textwidth]{Figures/conflict_gap05.eps}%  
%    \hfill
%    \includegraphics[width=0.50\textwidth]{Figures/conflict_gap_lpld05.eps}
%    \caption{latency tolerance, $\tau_{k}=0.5$ ms}
%\end{subfigure}
%\begin{subfigure}[b]{\linewidth}
%    \centering
%    \includegraphics[width=0.50\linewidth]{Figures/conflict_gap1.eps}%  
%    \hfill
%    \includegraphics[width=0.50\textwidth]{Figures/conflict_gap_lpld1.eps}
%    \caption{latency tolerance, $\tau_{k}=1$ ms}
%\end{subfigure}
%\begin{subfigure}[b]{\linewidth}
%    \centering
%    \includegraphics[width=0.50\linewidth]{Figures/conflict_gap2.eps}%  
%    \hfill
%    \includegraphics[width=0.50\textwidth]{Figures/conflict_gap_lpld2.eps}
%    \caption{latency tolerance, $\tau_{k}=2$ ms}
%\end{subfigure}
%\caption{Optimality gaps of the baseline heuristic \cite{you2018resource} and the variations of the conflict-aware heuristic CA (second column depict the LP-LD utilities) for $\tau_{k}=\{0.5, \text{ }1, \text{ }2\}$ ms, against the global optimum of P0. The $y$-label express the relative deviation to the optimum, expressed as percentage.}
%    \label{fig:opt_gap_1}
%\end{figure}

\begin{figure*}[h]
\centering
\begin{subfigure}[b]{\linewidth}
    \centering
    \includegraphics[width=0.33\textwidth]{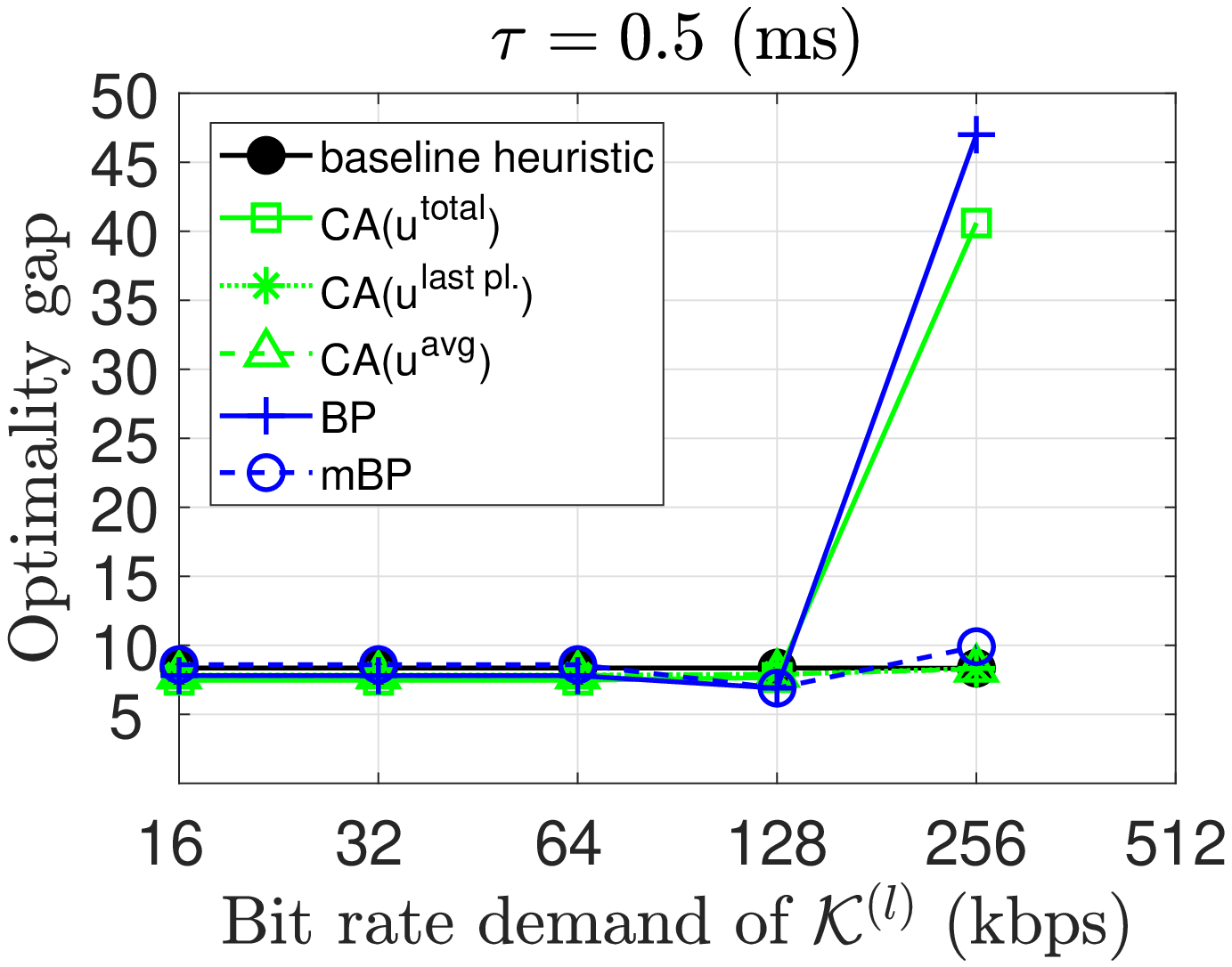}%
    \hfill
    \includegraphics[width=0.33\textwidth]{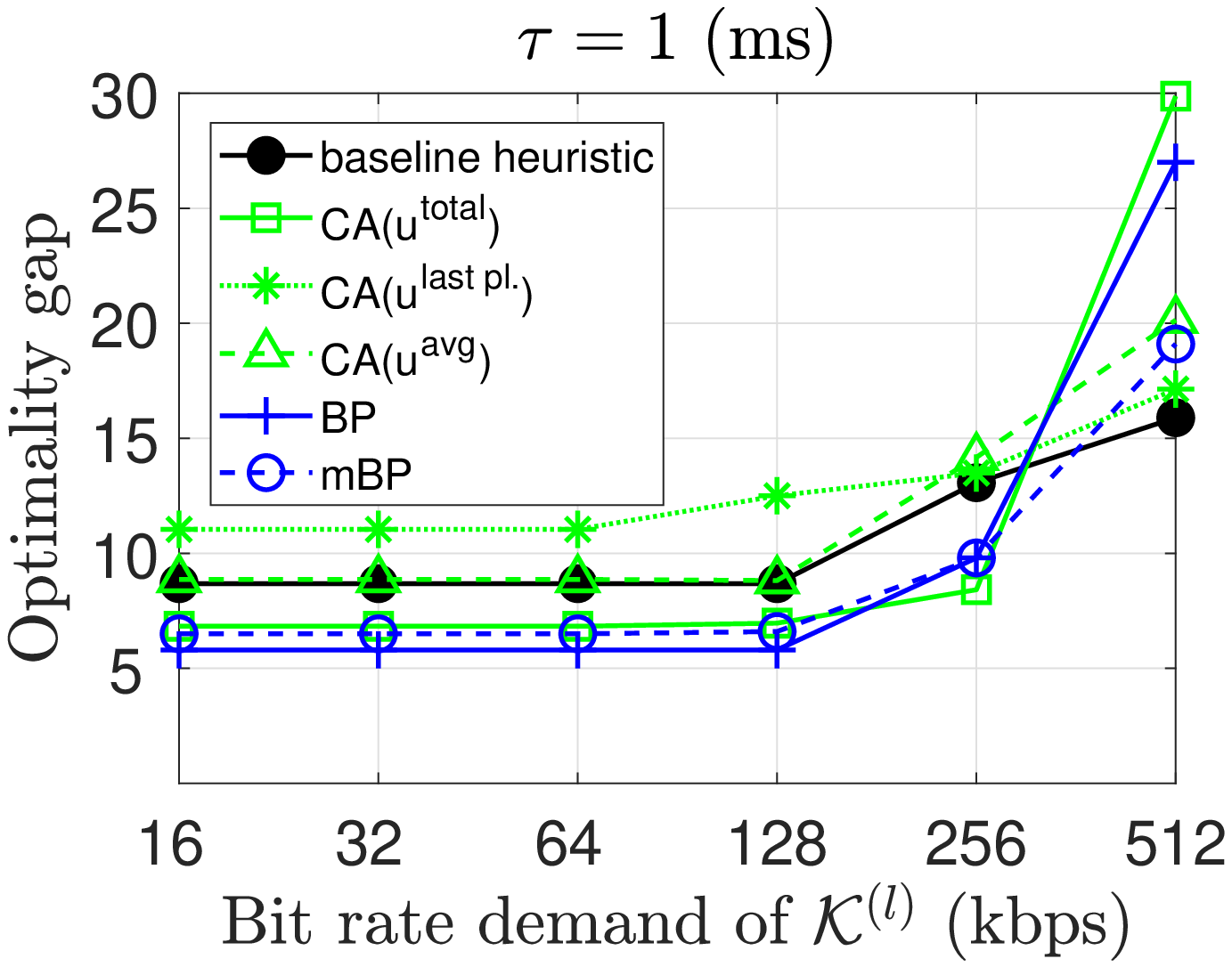}%
    \hfill
    \includegraphics[width=0.33\textwidth]{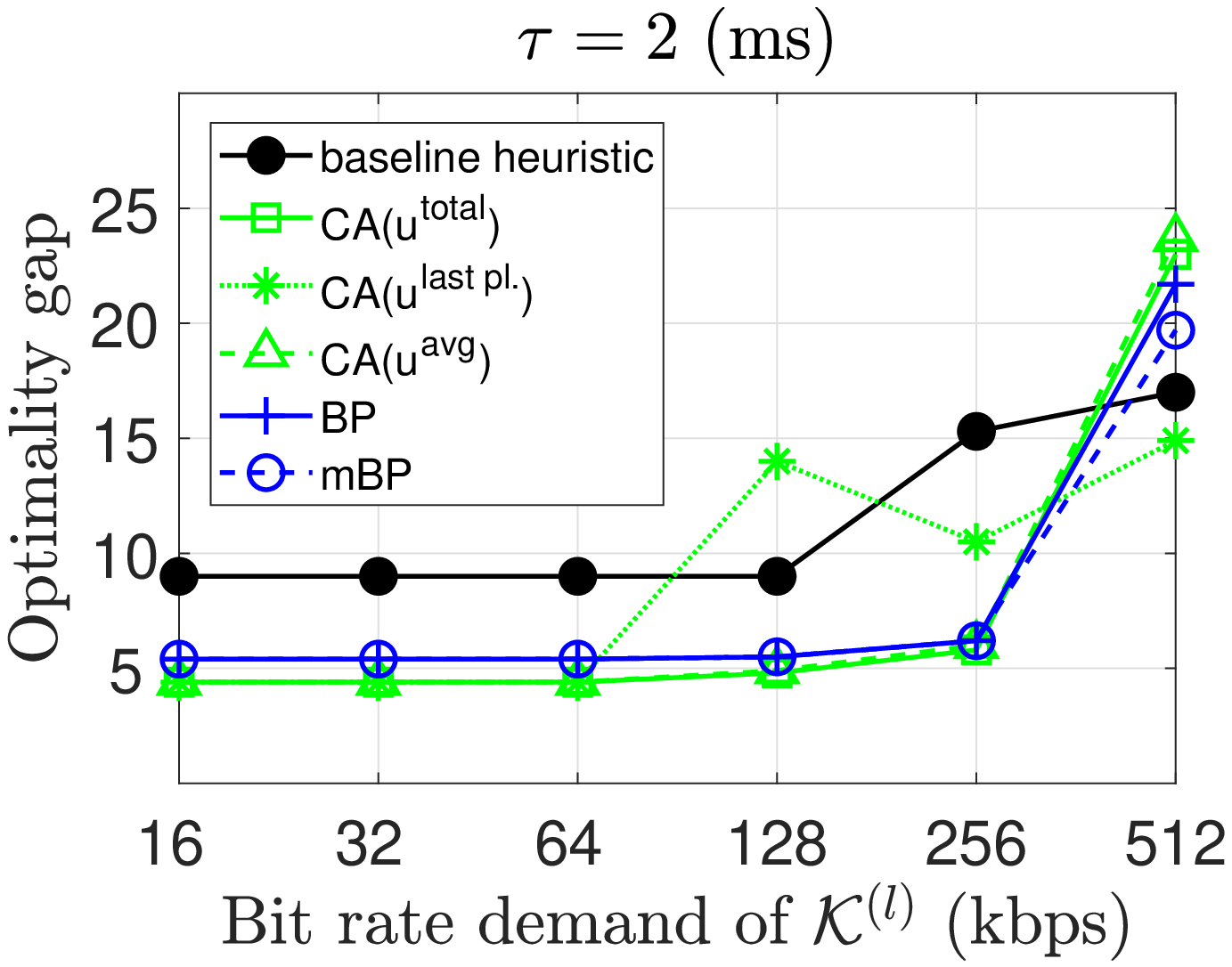}%
    \caption{Baseline heuristic and $CA$ heuristics.}
\end{subfigure}
\begin{subfigure}[b]{\linewidth}
\centering
    \includegraphics[width=0.33\textwidth]{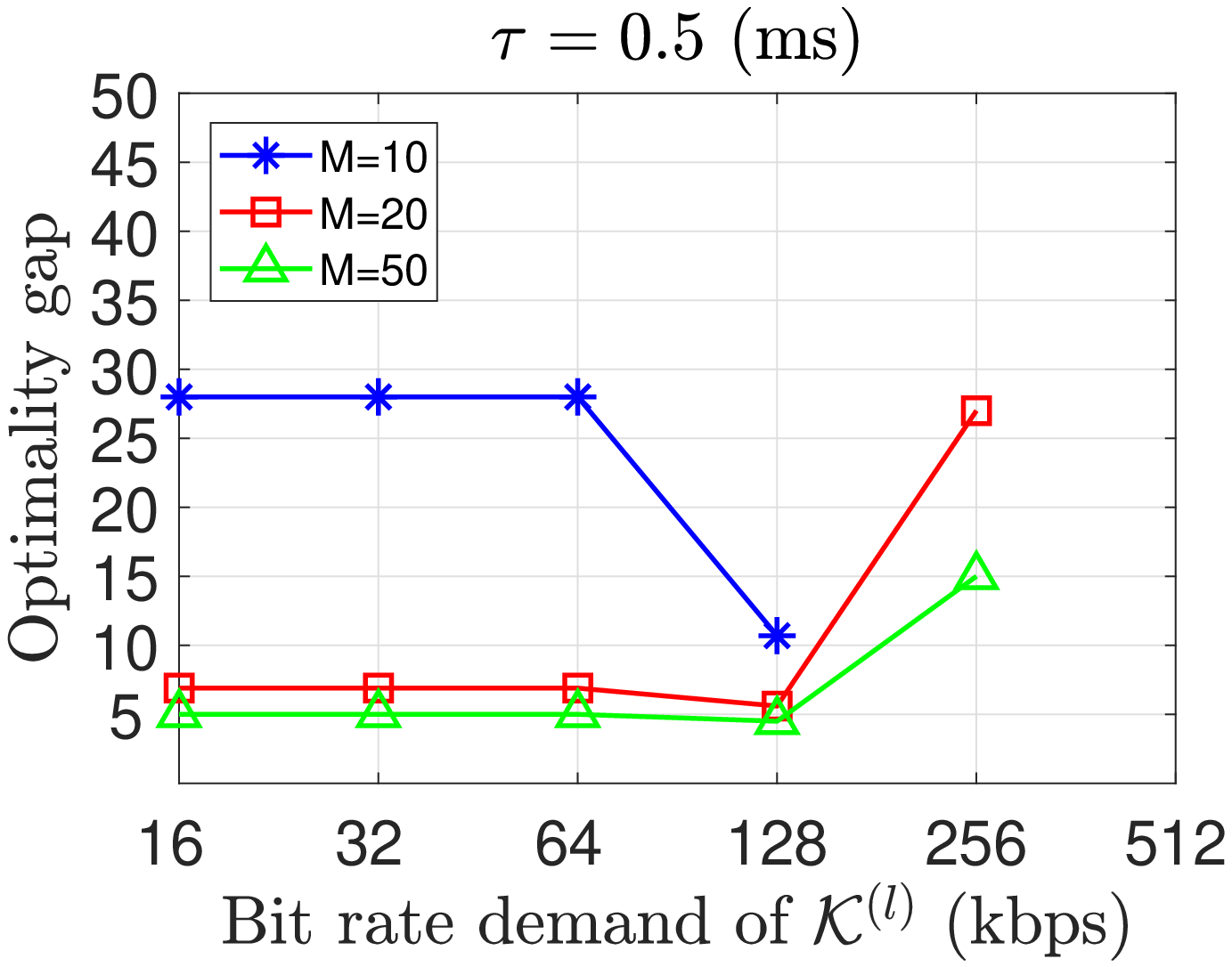}%
    \hfill
    \includegraphics[width=0.33\textwidth]{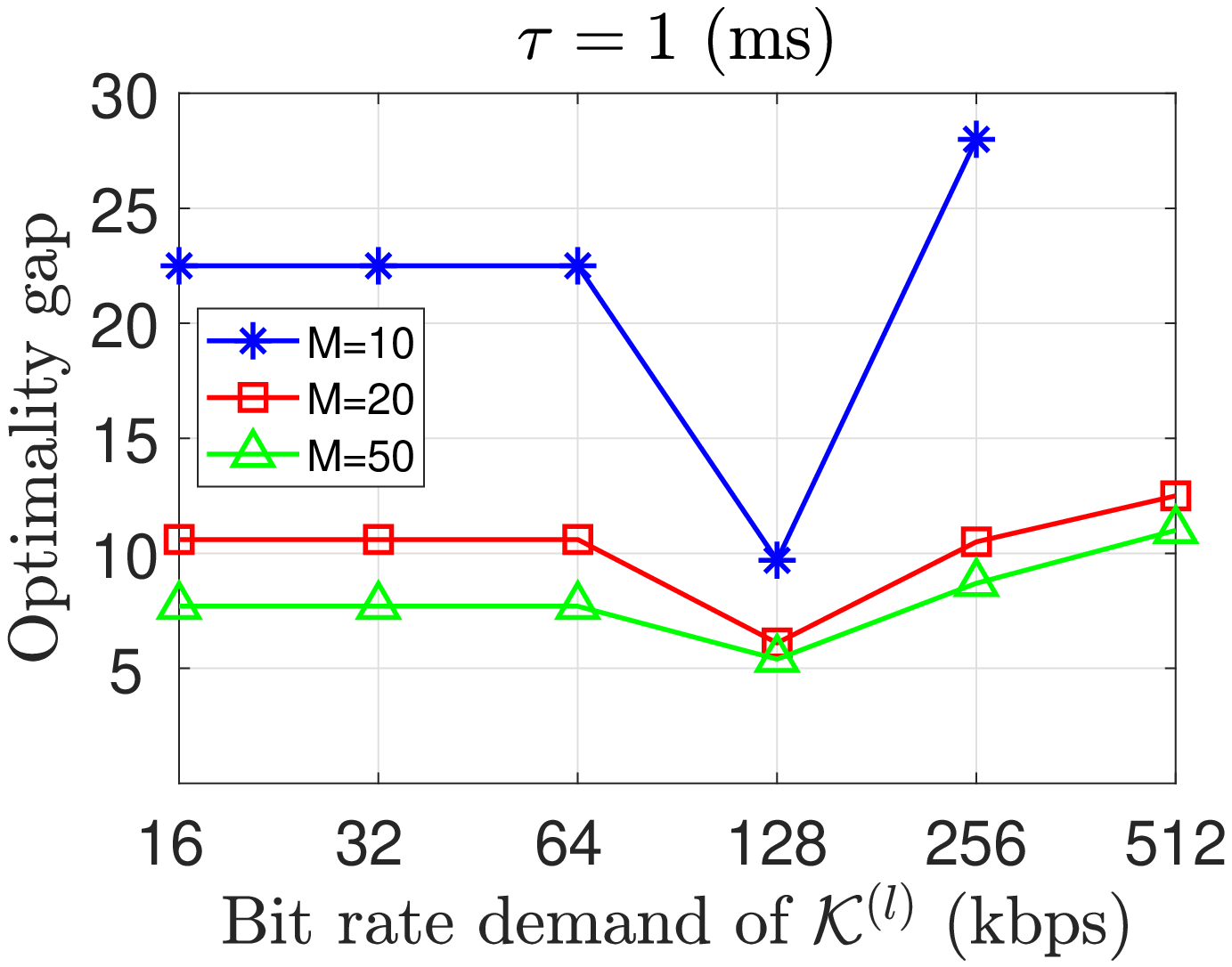}%
    \hfill
    \includegraphics[width=0.33\textwidth]{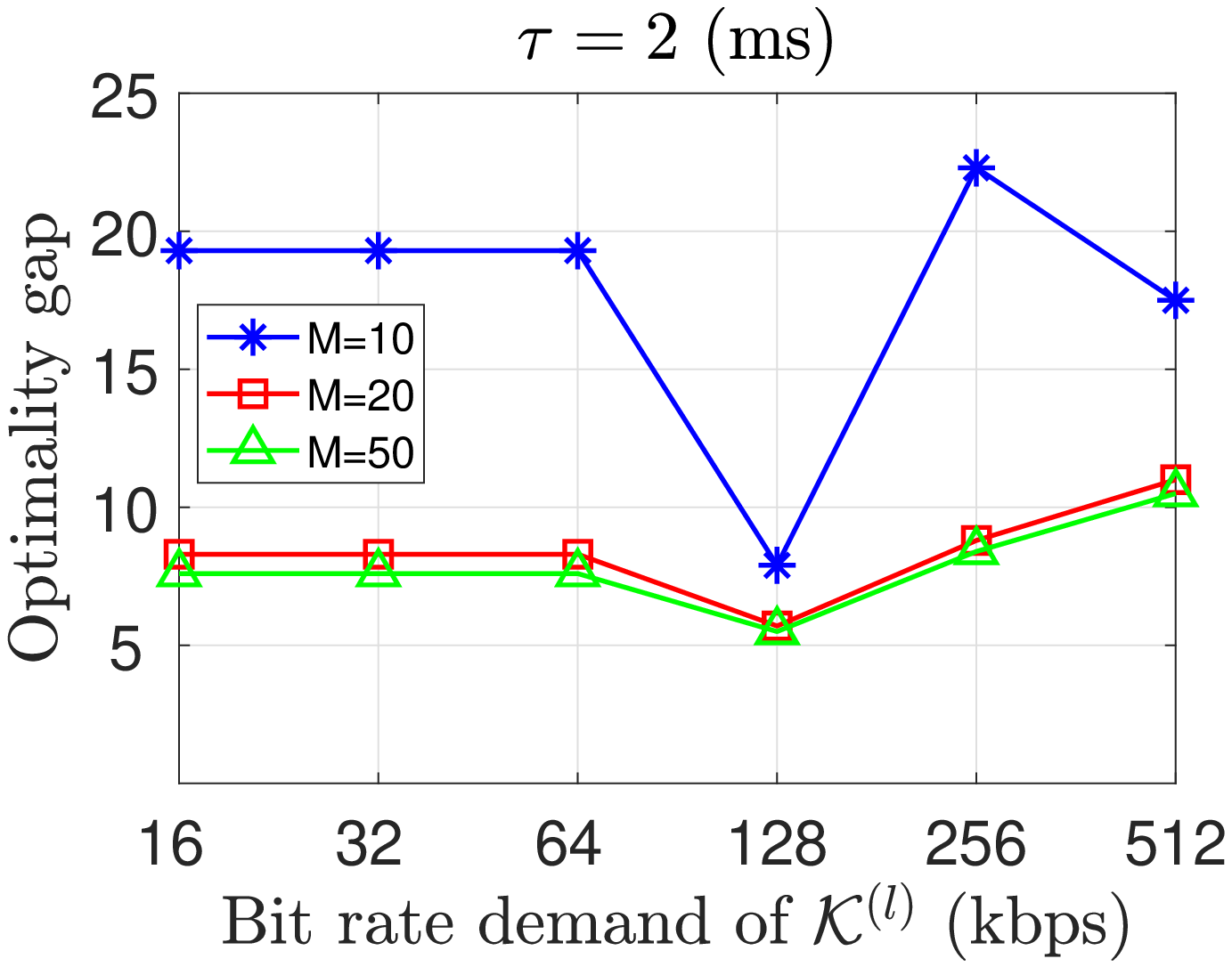}%
    %\caption{Baseline and $CA$ heuristics, using the LP-LD utility matrices.}
    \caption{Baseline heuristic, using the LP-LD utility matrices, for several thresholds $M$.}
\end{subfigure}
\caption{a) Optimality gaps: a) of the baseline heuristic \cite{you2018resource} and the variations of the conflict-aware heuristic CA, and, b) of the baseline LP-LD heuristic and thresholds for the sub-gradient iterations $M=\{10,\text{ } 20, \text{ }50\}$. Against the global optimum of P0, for latency tolerance values $\tau_{k}=\{0.5, \text{ }1, \text{ }2\}$ ms. The $y$-label express the relative deviation to the optimum, expressed as percentage.}
    \label{fig:opt_gap_1}
\end{figure*}

\begin{figure}[!t]
 \centering 
\includegraphics[width=0.5\textwidth]{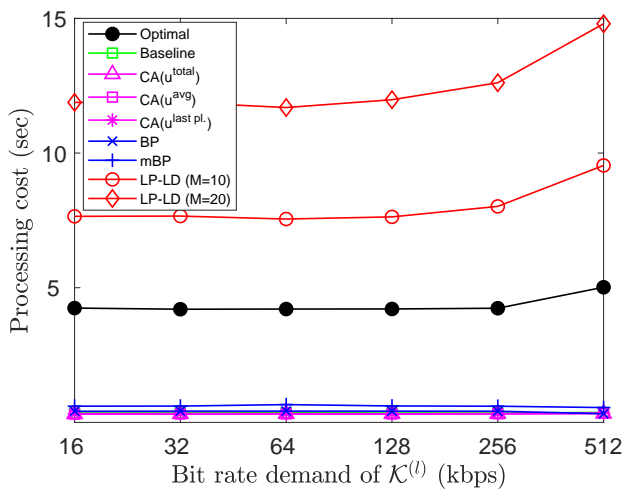}
\caption{The processing cost of: i) the optimal, ii) the baseline heuristic variations, iii) the bin packing based approach, and, iv) the LP-LD ($M=20$), for $\tau=1$ msec and $q_k=\{16, \text{ }32, \text{ }64, \text{ }128, \text{ }258, \text{ }512\}$ (kbps).}  
\label{costos} 
\end{figure}

Fig. \ref{fig:opt_gap_1} depicts the optimality gap: i) of the baseline, the variations of the conflict-aware and the bin packing based approaches (first row), and, ii) of the LP-LD relaxation of P0  (second row), for several values of maximum sub-gradient iterations, with respect to the bit rate demand and the latency tolerance of the $\mathcal{K}^{(l)}$ services. 

In the first row of Fig. \ref{fig:opt_gap_1} the conflict-aware and bin packing based heuristics are shown, in most cases, to outperform the baseline heuristic approach for bit rate demands up to $256$ kbps and higher latency tolerance values, see Figs. \ref{fig:opt_gap_1}(b) and (c), and to provide similar results for lower latency tolerance values, Fig. \ref{fig:opt_gap_1}(a). More precisely, $CA(\mathbf{u}^{total})$, $BP$ and $mBP$ clearly outperform all the other approaches, maintaining an optimality gap below to $10\%$ for $\tau_{k}=0.5$ msec and close to $5\%$ for $\tau_{k}=\{1,2\}$ msec. However, for the high bit rate demands of $512$ kbps, the above heuristics reduce their performance, due to the higher number of infeasible solutions; they do not satisfy the demands of the $k\in\mathcal{K^{(\ell)}}$ services. In such cases, $CA(\mathbf{u}^{last pl.})$ provide a superior performance, since a a more balanced (conservative) policy, like ${CA}(\mathbf{u}^{last \text{ }pl.})$, seems more suitable. As the $mBP$ algorithm also reduces the optimality gap when compared to the baseline $BP$ approach,  $mBP$ emerges  as an appropriate choice for low or or high bit rate demands of URLLC users.

The second row of Fig. \ref{fig:opt_gap_1} depicts the optimality gap of the LP-LD heuristic solutions, for various threshold values $M=\{10, 20, 50\}$ for the maximum sub-gradient iterations, against the global optimum. We do not provide the solutions coming from the incorporation of the utility matrices $\mathbf{u}_{LP},\text{ }\mathbf{u}_{LD}\in{\mathbb{R}_{\mathcal{B}\times{\mathcal{K}}}}$, since all variations conclude in similar results. 

As it is expected, higher threshold values of $M$ lead to a further reduction of the optimality gap, at the cost of a higher computational time. The choice of $M={10}$ results on very high optimality gaps, near to $20\%$ in most cases.  On the other hand for $M=20$ and $M=50$ the heuristics are shown to maintain the optimality gap close to $10\%$ even for $q= {512}$ kbps, except for $q= {256}$ kbps and $\tau= {0.5}$ msec. Note that the optimality gaps of the $CA(\mathbf{u}^{total})$ and the bin packing based approaches are slightly lower from that of the LP-LD variations for low throughput demands, as it can be seen by comparing the two rows of Fig. \ref{fig:opt_gap_1}. On the other hand, the reduction of the optimality gap using LP-LD utility matrices comes with a significant increase of the computational time, motivating further the use of the heuristic proposed in Section III.

Furthermore, we utilize our implementation to quantify the performance of the optimal and the heuristic approaches, in terms of processing cost. The computational time is measured on a Lenovo IdeaPad 510-15IKB laptop, with an Intel Core i7-7500U @ 2.70 GHz processor and 12 GB RAM. In Fig. \ref{costos}, we depict the processing cost of: i) the optimal solution, ii) the baseline heuristic variations (without the usage of the LP-LD utilities), iii) the bin packing based approach, and, iv) the LP-LD heuristic with threshold value $M=\{10, \text{ }20\}$, for  $q_k=\{16, 32, 64, 128, 256, 512\}$ and a conventional latency tolerance value $\tau=1$ ms. As it is depicted, the LP-LD solution is much more computational intensive than other heuristic approaches, even from the optimal solution. Note that higher threshold values increase drastically the processing cost, e.g., for $M=50$ the processing cost is of 22 sec. On the other hand, the processing cost of the bin packing and the conflict-aware heuristics is between 0.3 and 0.50 sec, indicating their low computational nature; we remind that the complexity of the conflict-aware and bin packing based heuristics is of $\mathcal{O}=N\log(N)$.

\begin{figure}
 \centering 
\includegraphics[width=0.5\textwidth]{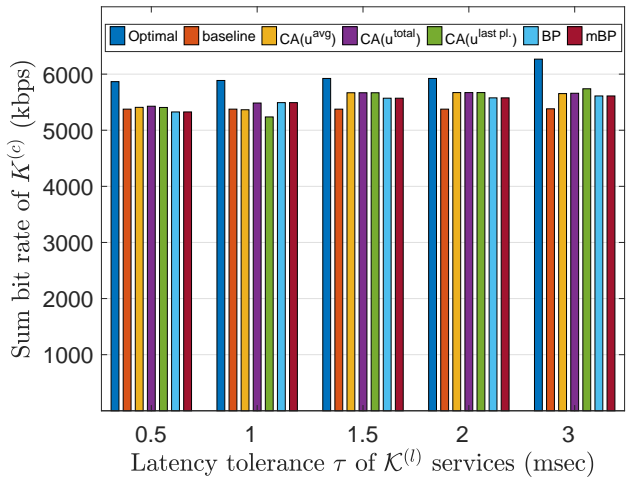}
\caption{Sum bit rate of $\mathcal{K}^{(c)}$ services for various values for the latency tolerance of $\mathcal{K}^{(\ell)}$ services, when the bit rate demands of $\mathcal{K}^{(\ell)}$ users are all equal and set to $64$ kbps. Similar results are produced for demands of $16$ and $32$ kbps.}  
\label{fig:64kbps} 
\end{figure}

\begin{figure}
\centering
\includegraphics[width=0.5\textwidth]{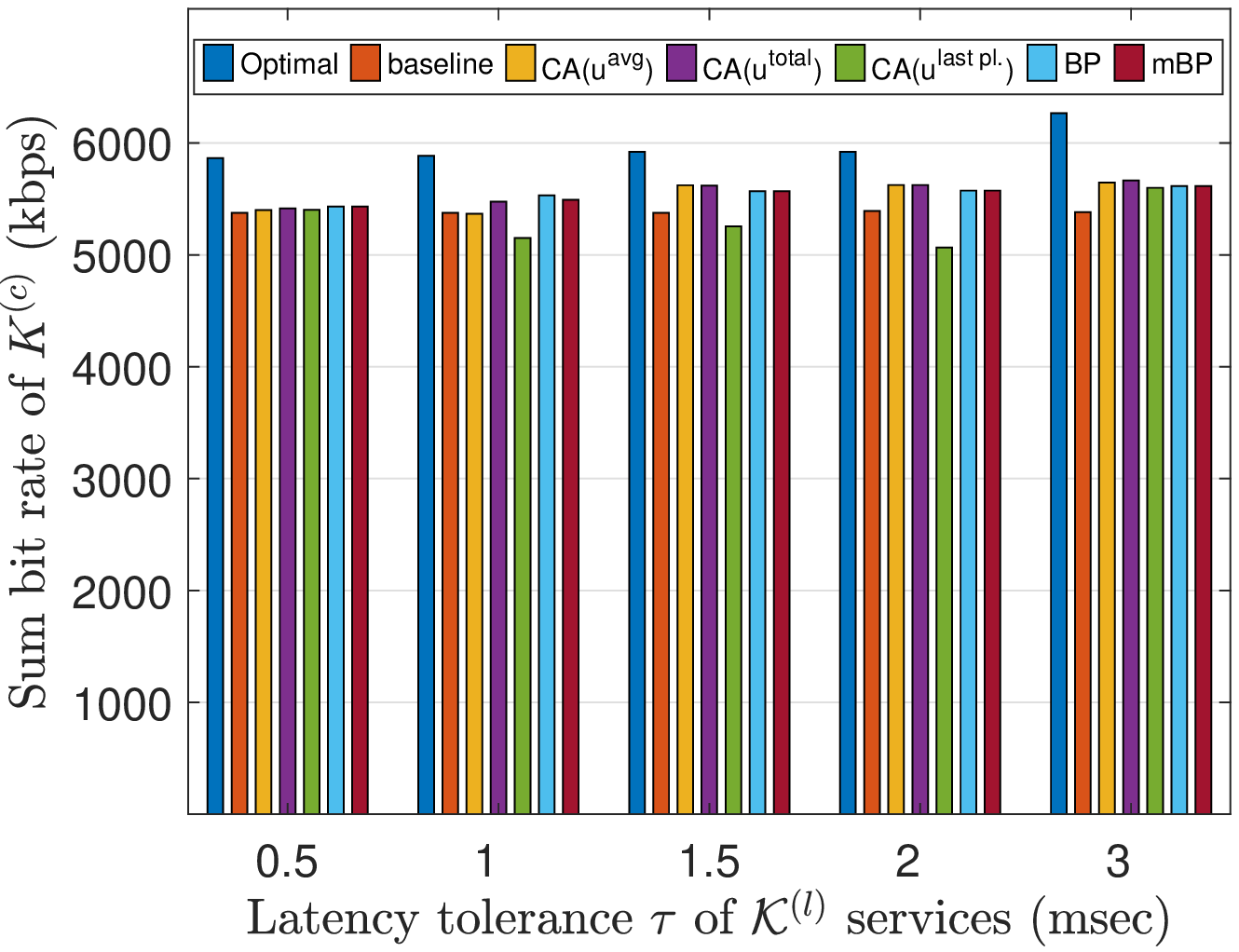}
\caption{Sum bit rate of $\mathcal{K}^{(c)}$ services for various values for the latency tolerance of $\mathcal{K}^{(\ell)}$ services, when the bit rate demands of $\mathcal{K}^{(\ell)}$ users are all equal and set to $128$ kbps.} 
\label{fig:128kbps}
\end{figure}

\begin{figure}
\centering
\includegraphics[width=0.5\textwidth]{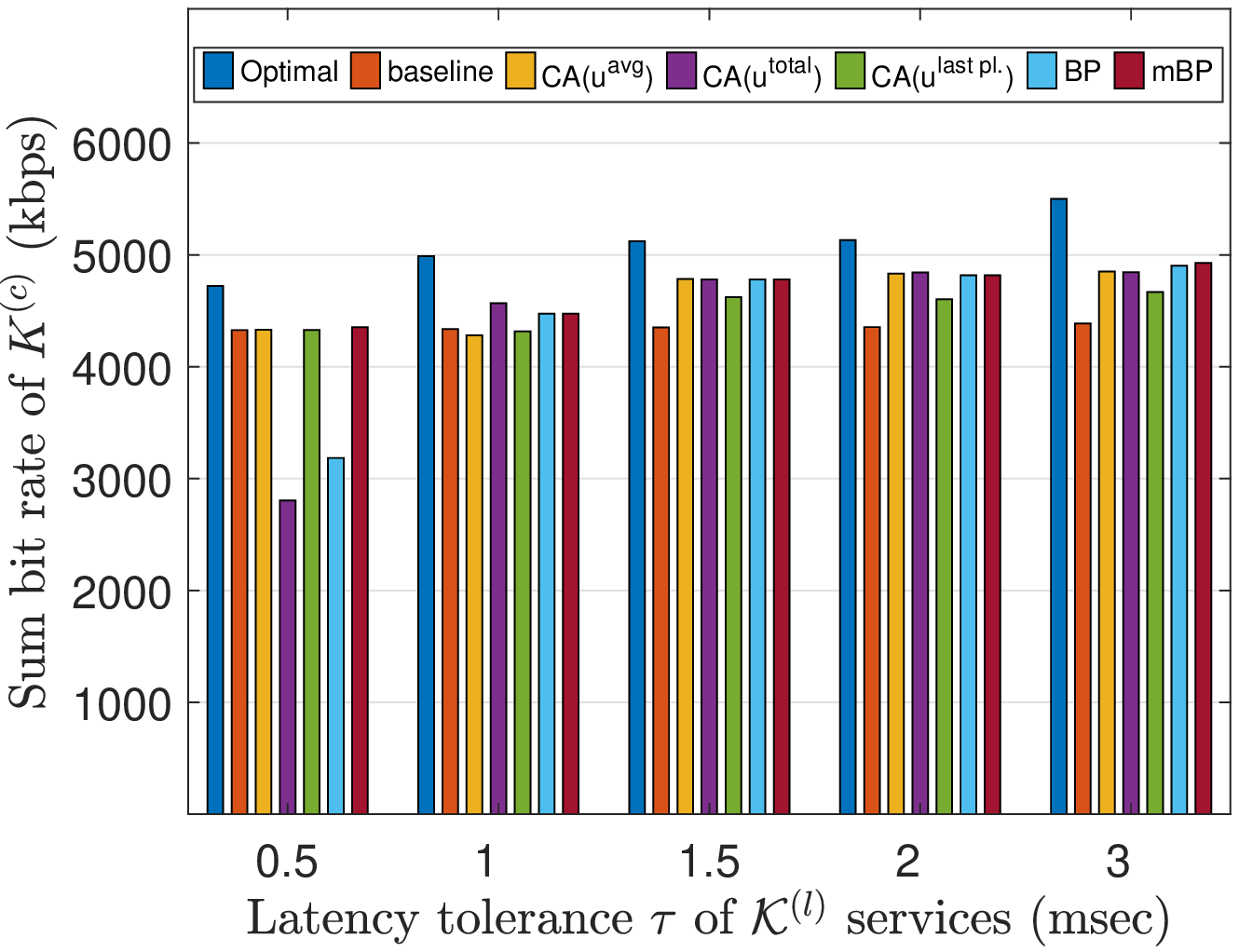}
\caption{Sum bit rate of $\mathcal{K}^{(c)}$ services for various values for the latency tolerance of $\mathcal{K}^{(\ell)}$ services, when the bit rate demands of $\mathcal{K}^{(\ell)}$ users are all equal and set to $256$ kbps.} 
\label{fig:256kbps}
\end{figure}

\begin{figure}
\centering
\includegraphics[width=0.5\textwidth]{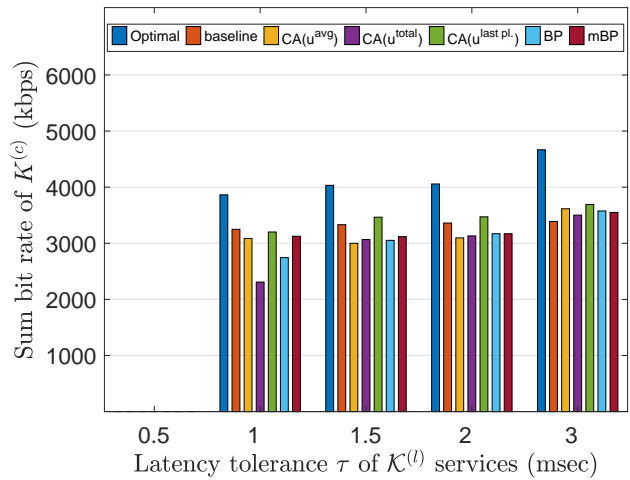}
\caption{Sum bit rate of $\mathcal{K}^{(c)}$ services for various values for the latency tolerance of $\mathcal{K}^{(\ell)}$ services, when the bit rate demands of $\mathcal{K}^{(\ell)}$ users are all equal and set to $512$ kbps.} 
\label{fig:512kbps}
\end{figure}

%\begin{figure}[!t]
%\centering
%\includegraphics[width=1.00\columnwidth]{Figures/1024.jpg}
%\includegraphics[width=0.5\textwidth]{Figures/q1024n.eps}
%\caption{Sum bit rate of $\mathcal{K}^{(c)}$ services for various values for the latency tolerance of $\mathcal{K}^{(\ell)}$ services, when the bit rate demands of $\mathcal{K}^{(\ell)}$ users are all equal and set to $1024$ kbps.} 
%\label{fig:1024kbps}
%\end{figure}

% \begin{figure}[!t]
% \centering
% \vspace{-0.0em}
% \includegraphics[width=1.00\columnwidth]{Figures/opt_gap.jpg}
% %\vspace{-0.8em}
% \caption{The average relative deviation of obtained solutions to the optimum} 
% \label{fig:opt_gap}
% \vspace{-0.0em}
% \end{figure}

Next, we compare the performance of the conflict-aware heuristic solutions (Algorithm 1), the heuristic algorithms inspired from the reformulation of the scheduling problem as a bin packing optimization (Algorithm 2), the baseline heuristic and the optimal solution. We exclude the $CA$ approaches based on the LP-LD utility matrices from the comparison, as these come at the cost of a significantly higher complexity. Fig. \ref{fig:64kbps} depicts a comparable performance of the heuristic algorithms to the global optimum (obtained through Gurobi solvers), while keeping the complexity very low. Note that the proposed algorithms, exceed the performance of the baseline heuristic, especially for $\tau>0.5$ msec. This showcases that indeed, the reformulation of the optimal scheduling as a conflict minimization problem is highly pertinent and allows shedding light on how to jointly address the constraints (\ref{eq:constraint1}) and (\ref{eq:constraint2}) of P0. It is also noteworthy that more elaborate heuristics could be proposed in the same context, by looking at algorithms with lower optimality gaps to the optimal bin packing solution. 

The same conclusions can be reached in Figs. \ref{fig:128kbps} and \ref{fig:256kbps} for URLLC demands of $128$ and $256$ kbps, respectively. In these cases, all the conflict-aware choices exceed the performance of the baseline heuristic; the choice of $CA(\mathbf{u}^{last pl.})$ metric is the only one with lower performance to that of baseline heuristic for $q=128$ kbps. We remind that the instabilities of the $BP$ and the $CA(\mathbf{u}^{total})$ approaches, for $\tau=0.5$ msec and $q=256$ kbps, is due to the higher amount of infeasible solutions, which is not the case in the other heuristic solutions including the mBP approach. 

Moreover, in case of higher bit rate demands for the URRLC users, specifically for $q=512$ kbps (Fig. \ref{fig:512kbps}), the bin packing based approach seems to exceed the performance of $CA(\mathbf{u}^{total})$ and $CA(\mathbf{u}^{avg})$ providing a performance close to that of the $CA(\mathbf{u}^{last pl.})$ heuristic, where the latter exceed the performance of the baseline heuristic for the chosen latency tolerance values. Furthermore, the proposed heuristics results overpass the performance of the baseline heuristic for high latency tolerance values, $\tau=3$ msec. 

Finally, the conflict-aware heuristic based on the variations $CA(\mathbf{u}^{avg}$ and $CA(\mathbf{u}^{avg}$, and the bin packing based solutions verify their superior performance to that of the baseline heuristic for $q\leqslant{256}$ kbps, as it is also shown in Fig. \ref{fig:opt_gap_1}.

%% file: Sections/conclusion.tex
In 5G and beyond networks, URLLC services will coexist with eMBB services through challenging layer 2 scheduling. To address the latter, we have reformulated the standard eMBB throughput maximization problem as an equivalent conflict minimization, which points at minimizing the overall amount of conflicts.
Building on this premise, two lightweight and efficient scheduling approaches were proposed: a family of conflict-aware heuristics that employ conflict aware utilities and a heuristic inspired by the bin packing problem. 

In addition to the proposed scheduling using orthogonal multiple access (OMA),
we further proposed the use of non-orthogonal multiple access (NOMA) to mitigate conflicts. We investigated the potential advantages of allowing for non-orthogonal sharing of radio resources with flexible numerology and frame structure. The intuition for NOMA's superior performance, as a result of alleviating conflicts, was demonstrated to hold; importantly, NOMA can potentially offer significant advantages particularly in the case of ultra-low latency constraints for the URLLC users.

Extensive simulations were performed for URLLC services with different QoS requirements both for OMA and NOMA scenarios. The simulation results showed that i)  all of the proposed heuristics have near-optimal performance, demonstrating that conflict minimization is indeed key to layer 2 scheduling, and, ii) there are significant gains in terms of resource utilization when employing NOMA.